\documentclass[a4paper,fleqn,usenatbib,useAMS]{article}

\usepackage{graphicx}	
\usepackage{multicol}        
\usepackage{pdflscape}	
\RequirePackage{bm,amsfonts,hhline,amsmath,scrextend,amssymb,microtype,float,eurosym,latexsym,epsf,mathtools,cuted,times,makecell,array,natbib,titling}
\usepackage[font={footnotesize,it}]{caption}
\usepackage[caption=false]{subfig}
\usepackage{epstopdf}
\usepackage[font={footnotesize,it}]{caption}
\usepackage[switch]{lineno}
\usepackage{hyperref}

\begin{document}

\title{Anisotropic compact stars in $f(T)$ gravity under Karmarkar condition}

\author{Debabrata Deb$^{a,1}$, Shounak Ghosh$^{a,2}$, S.K. Maurya$^{b,3}$, Maxim Khlopov$^{c,d,4}$  \and Saibal Ray$^{e,f,5}$}
\date{%
    $^a$Department of Physics, Indian Institute of Engineering Science and Technology, Shibpur, Howrah, West Bengal, 711103, India\\%
    $^b$Department of Mathematical and Physical Sciences, College of Arts and Science, University of Nizwa, Nizwa, Sultanate of Oman\\
    $^c$National Research Nuclear University, “MEPHI” (Moscow Engineering Physics Institute), Moscow 115409, Russia\\
    $^d$ APC Laboratory, 10 rue Alice Domon et L{\'e}onie Duquet, 75205 Paris Cedex 13, France\\
    $^e$Department of Physics, Government College of Engineering and Ceramic Technology, Kolkata 700010, West Bengal, India\\
    $^f$Department of Natural Sciences, Maulana Abul Kalam Azad University of Technology, Haringhata 741249, West Bengal, India\\[2ex]%
}

\maketitle

\footnotetext[1]{ddeb.rs2016@physics.iiests.ac.in}
\footnotetext[2]{shnkghosh122@gmail.com}
\footnotetext[3]{sunil@unizwa.edu.om}
\footnotetext[4]{khlopov@apc.in2p3.fr}
\footnotetext[5]{saibal@associates.iucaa.in}

\clearpage


\title{Anisotropic compact stars in $f(T)$ gravity under Karmarkar condition}
\author{~~}
\date{~~}
\maketitle

\begingroup
\fontsize{14pt}{12pt}\selectfont
\noindent {\bf Abstract}
\vspace{0.3cm}
\endgroup \\
	In this study we present a generalized spherically symmetric, anisotropic and static compact stellar model in $f(T)$ gravity, where $T$ represents the torsion scalar. By employing the Karmarkar condition we have obtained embedding class 1 metric from the general spherically metric of class 2 and the solutions of the Einstein field equations (EFE) has been presented with the choice of suitable parametric values of $n$ under a simplified linear form of $f(T)$ gravity reads as $f(T)=A+BT$, where $A$ and $B$ are two constants. By matching the interior spacetime metric with the exterior Schwarzschild metric at the surface and considering the values of mass and radius of the compact stars we obtain the values of the unknown constants. We have presented further a detailed analysis of the physical acceptability and examined the stability of the stellar configuration by studying the energy conditions, generalized Tolman-Oppenheimer-Volkov (TOV) equation, Herrera cracking concept, adiabatic index, etc. In the investigation, we predict numerical values of the central density, surface density, central pressure, etc., in a tabular form taking different values of $n$ specifically for $LMC~X-4$, $Cen~X-3$ and $SMC~X-1$ as the representative of compact star candidates. \\	 


\section{Introduction}\label{sec:intro}
 In the past few decades, the local isometric embedding of general relativistic Riemannian manifolds in pseudo-Euclidean space of dimension $m\geq4$ has drawn attention of many researchers.~\citet{Schlai1871} first addressed the embedding problem on geometrically significant spacetimes. Later on~\citet{Nash1956} presented the isometric embedding theorem. Interestingly, the unification of gravity with the particle interactions can be achieved by embedding 4-dimensional EFE of general relativity (GR) into a higher dimensional space, like Kaluza-Klein model~\citep{Kaluza1921,Klein1926}. Recently Tavakol and coworkers~\citep{Tavakol1,Tavakol2,Tavakol3} by employing Campbell's theorem~\citep{Campbell1926} have presented a relation between manifolds which differs by one dimension like a ladder. The lower end of the ladder represents the embedding of $4D$ Einstein equations into the $5D$ Ricci-flat equations by setting the Ricci tensor to zero. In particular, 5th-dimension is the simplest extension of the spacetime. Also, the fifth dimension is considered as the low-energy limit of higher dimensional theories of particles, viz., 11$D$ supergravity, 10$D$ supersymmetry and higher-$D$ versions of string theory, etc.

\citet{Schlaefli1871} conjectured that the neighborhood in a $n$-dimensional manifold requires an embedding space of $m\left[= n(n+1)/2\right]$ dimensional pseudo-Euclidean space. It is referred as class $m$ if $n$-dimensional Riemannian space is embedded into lowest $n+m$ dimensional space. The embedding class of the spherically symmetric and plane symmetric spacetime are 2 and 3 respectively, whereas the Standard Model of modern cosmology, i.e., the Friedmann-Lemaitre-Robertson-Walker spacetime~\citep{F1,L1,R1,W1} and the interior Schwarzschild solution is of embedding class 1. On the other hand, the \citet{Schwarzschild1937} exterior solution is of embedding class 2 and the embedding class of the Kerr metric is 5~\citep{Kuzeev1980}. In the present study we are concerned with the embedding spherically symmetric static metric into the 5-$D$ Euclidean space, i.e, embedding class 1 metric. 

In the framework of GR the effect of anisotropy on the spherically symmetric stellar objects have been studied explicitly in literature (vide review of~\citet{HS1997}, and also~\citet{MH2002} and references therein). Existence of superfulid, phase transition, the presence of the magnetic field or external field, rotational motion, the presence of fluids of different types, etc. are the reasons for the anisotropy in a relativistic stellar system. In the present study of the relativistic stellar body, as the system is static, hence the anisotropic stress may be originated due to the presence of anisotropic fluid or the elastic nature of the super fluid or presence of the superfluid. In this line several works~\citep{BL1974,MH2002,Kalam2005,Horvat2011,Dev2003} can be noted.~\citet{Ruderman1972} in his pioneer work on the astrophysical anisotropy predicted that in the high density ($>{10}^{15}$~$gm/cm^3$) nuclear matter interacts relativistically and exhibits anisotropy as the inherent property.~\citet{BL1974} argued that in the presence of complex strong interactions superfluidity and superconductivity are the reasons for anisotropy inside the superdense matter. Note that raise of anisotropy in the pressure components is relevant for the diverse situations, such as when the radial part of the pressure, ${p_r}(r)$, is different from the angular part, ${p_{\theta}}(r) = {p_{\phi}}(r) \equiv {p_{t}}(r)$, it is defined as anisotropic pressure, where ${p_{\theta}}(r) = {p_{\phi}}(r)$ and both of the components are functions of the radial coordinate as a direct consequence of spherical symmetry. In a physical system, the pressure is anisotropic when the scalar field has non-zero spatial gradient. Also, the origin of local anisotropy in self gravitating system and its effects on the system are studied in detail by~\citet{HS1997}. Later, many authors have investigated effect of the local anisotropy on the spherically symmetric static body~\citep{Ivanov2002,SM2003,MH2003,Usov2004,Varela2010,Rahaman2010,Rahaman2011,Rahaman2012,Kalam2012,Deb2015,Shee2016,Maurya2016,Maurya2017}
.

The concept of parallelism, an equivalent formation of GR, is receiving attention for the past few years as an alternative theory of gravity and well known as the teleparallel equivalent of GR (TEGR)~\citep{Mollar1961,Pellegrini1963,Hayashi1979}. The idea behind this formalism is the consideration of more general manifold which consider besides the curvature a quantity called torsion.~\citet{Ferraro1,Ferraro2} in their work studied the modifications of TEGR in the context of cosmology which is known as $f(T)$ gravity theory. The appealing part of $f(T)$ gravity is that its field equations are of second order unlike $f(R)$ gravity and it is constructed with a generalized Lagrangian~\citep{Bengochea2009,Linder2010}.

In the field of cosmology, whether theoretical or observational investigations, several researchers successfully used the $f(T)$ gravity in their study~\citep{Wu2010a,Tsyba2011,Dent2011,Chen2011,Bengochea2011,Wu2010b,Yang2011,Zhang2011,Li2011,Wu2011,Bamba2011}. \citet{Deliduman2011} and also \citet{Wang2011} implemented $f(T)$ gravity to find static and spherical symmetric solution for the relativistic stars.~\citet{Deliduman2011} based on the general relativistic conservation equation demanded that relativistic stars in $f(T)$ gravity do not exist, however~\citet{Boehmer2011} showed in their study that they do exist. In this line other works on $f(T)$ gravity can be consulted for further study~\citep{Andrade2000,Ferraro2011,Tamanini2012a,Tamanini2012b,Aldrovandi2013,Aftergood2014,Abbas2015a,Abbas2015b}.

In the present article we have tried to investigate compact stellar system under the embedding class 1 metric in the framework of $f(T)$ gravity by exploiting the results presented by \citet{Boehmer2011}. We demonstrated our calculations considering, specifically the following three stars $LMC~X-4$, $Cen~X-3$~and~$SMC~X-1$, as the representative of compact stars with the suitable choice of the parametric values of $n$. Against the above background we execute our work in the following scheme: the basic mathematical formulations of $f(T)$ gravity are shown in Sec.~\ref{sec 2}. The EFE under $f(T)$ gravity have been provided in Sec.~\ref{sec 3}. In Sec.~\ref{sec4} class one condition is implemented and the general solutions are achieved. The constants are determined in Sect.~\ref{sec5} with the help of boundary conditions. Some of the physical features of the model, especially the energy conditions and stability of the stellar system are studied in Sec.~\ref{sec6}. In Sec.~\ref{sec7} we investigate the compactification factor and the redshift of the compact star and finally we have concluded our study through the tabular presentation of the important results and discuss the physical validity in Sec.~\ref{sec8}.

 \section{Basic mathematical formalism using $f(T)$ gravity}\label{sec 2}
For the basic formulation of $f(T)$ gravity \citet{Boehmer2011} have used some Greek and Latin indices  correspond to the spacetime coordinates and tetrad fields, respectively. Following them the metric for the spacetime can be written as
\begin{equation}\label{2.1}
ds^{2}=g_{\mu\nu}dx^{\mu}dx^{\nu}.
\end{equation}

Use of the tetrad matrix can transform the above metric to Minkowskian as
\begin{eqnarray}\label{2.2}
dS^{2}=g_{\mu\nu}dx^{\mu}dx^{\nu}=\eta_{ij}\theta^{i}\theta^{j},\\ \label{2.3}
dx^{\mu}=e_{i}^{\mu}\theta^{i}, \theta^{i}=e^{i}_{{\mu}}dx^{\mu},
\end{eqnarray}
where $\eta_{ij}=diag[1,-1,-1,-1]$ and
$e_{i}^{\mu}e_{i}^{\nu}=\delta_{\nu}^{\mu}$ or
$e_{i}^{\mu}e_{j}^{\nu}=\delta_{i}^{j}$.

Here $\sqrt{-g}=det[e_{\mu}^{i}]=e$ is the root of the metric determinant. Now for the vanishing Riemann tensor part and non-vanishing torsion term the Weitzenbock's connection components are
\begin{equation}\label{2.4}
\Gamma^{\alpha}_{\mu\nu}=e_{i}^{\alpha}\partial_{\nu}e_{\mu}^{i}=-e_{i}^{\mu}\partial_{\nu}e_{i}^{\alpha}.
\end{equation}

So the torsion and the cotorsion are given as
\begin{eqnarray}\label{2.5}
T^{\alpha}_{\mu\nu}=\Gamma^{\alpha}_{\nu\mu}-\Gamma^{\alpha}_{\mu\nu}=e_{i}^{\alpha}(\partial_{\mu}e_{\nu}^{i}-\partial_{\nu}e_{\mu}^{i}), \\ \label{2.6}
K^{\mu\nu}_{\alpha}=-\frac{1}{2}(T^{\mu\nu}_{\alpha}-T^{\nu\mu}_{\alpha}-T^{\mu\nu}_{\alpha}).
\end{eqnarray}

Also the tensor $S_{\alpha}^{\mu\nu}$ can be written as
\begin{equation}\label{2.7}
S_{\alpha}^{\mu\nu}=\frac{1}{2}(K^{\mu\nu}_{\alpha}+\delta^{\mu}_{\alpha}T^{\beta\nu}_{\beta}-\delta^{\nu}_{\alpha}T^{\beta\mu}_{\beta}).
\end{equation}

Using torsion and cotorsion the torsion scalar is provided as
\begin{equation}\label{2.8}
T=T^{\alpha}_{\mu\nu}S_{\alpha}^{\mu\nu}.
\end{equation}

The action for $f(T)$ gravity is defined as (using natural units $G=c=1$)
\begin{equation}\label{2.9}
S[e_{\mu}^{i}, \Phi_A]=\int d^4 x
\left[\frac{1}{16\pi}f(T)+\mathcal{L}_{Matter}(\Phi_A)\right],
\end{equation}
where the $\mathcal{L}_{Matter}(\Phi_A)$ corresponds to the matter field and $f(T)$ represents any arbitrary analytic function of $T$. If one vary the above action with respect to tetrad, one have field equations due to $f(T)$ gravity as~\cite{Dent2011}
\begin{eqnarray}\label{2.10}
& \qquad S_{\mu}^{\nu\rho}\partial_{\rho}T
f_{TT}+[e^{-1}e^{i}_{\mu}\partial_{\rho}(ee_i^{\alpha}S_{\alpha}^{\nu\rho})+T^{\alpha}_{\lambda\mu}
S_{\alpha}^{\nu\lambda}]f_T +\frac{1}{4}\delta^{\nu}_{\mu}f=4\pi \mathcal{T}_{\mu}^{\nu},
\end{eqnarray}
where $\mathcal{T}_{\mu}^{\nu}$ is the energy-momentum tensor for anisotropic fluid given as
\begin{equation}\label{2.11}
\mathcal{T}_{\mu}^{\nu}=(\rho+p_{t})u_{\mu}u^{\nu}-p_{t}\delta_{\mu}^{\nu}+(p_{r}-p_{t})v_{\mu}v^{\nu},
\end{equation}
where $v^{\mu}$ and $u^{\mu}$ are the 4-radial vectors and 4-velocity vectors, respectively whereas $p_r$ and $p_t$ are the 
radial and tangential pressures, respectively.

\section{Basic field equations}\label{sec 3}
The interior spacetime of the compact star is assumed to be static and spherically symmetric and the line element can be described by the Schwarzschild metric as
\begin{equation}\label{3.1}
ds^{2}=e^{\nu(r)}dt^{2}-e^{\lambda(r)}dr^{2}-r^{2}(d\theta^{2}+\sin^{2}\theta d\phi^{2}).
\end{equation}
 Let us introduce the tetrad matrix for the metric (\ref{3.1}) as follows:
\begin{equation}\label{3.2}
[e_{\mu}^{i}]=diag[e^\frac{\nu(r)}{2},e^\frac{\lambda(r)}{2},r,r\sin\theta].
\end{equation}

Now using Eq. (\ref{3.2}), we have $e=det[e_{\mu}^{i}]=e^\frac{(\nu+\lambda)}{2}r^{2}\sin \theta$.
By employing Eqs. (\ref{2.4})-(\ref{2.8}), the torsion scalar and its derivative can be derived in terms of $r$ as
\begin{eqnarray}\label{3.3}
T(r)=\frac {2 e^{-\lambda}}{r}\left(\nu^{\prime}+\frac {1}{r}\right), \\\label{3.4}
T^{\prime}(r)=\frac {2 e^{-\lambda}}{r}\left[\nu^{\prime\prime}+\frac
{1}{r^{2}}-T \left(\lambda^{\prime}+\frac{1}{r}\right)\right],
\end{eqnarray}
where prime denotes derivative with respect to $r$.

Now substituting the above tetrad field (\ref{3.2})and inserting the values of torsion scalar and its derivative one can find out the Einstein field equations explicitly in $f(T)$ field as
\begin{eqnarray}\label{3.5}
& \qquad\hspace{-1cm} \frac {f}{4}-\frac{f_{T}}{2}\left[T-\frac{1}{r^{2}}-\frac {
e^{-\lambda}}{r}(\nu^{\prime}+\lambda^{\prime})\right]=4\pi\rho ,\\\label{3.6}
& \qquad\hspace{-1cm} \frac{f_{T}}{2}\left(T-\frac{1}{r^{2}}\right)-\frac
{f}{4}=4\pi p_{r},\\\label{3.7}
& \qquad\hspace{-1cm} \left[\frac{T}{2}+e^{-{\lambda}}\left\lbrace\frac{{\nu}^{\prime\prime}}{2}+\left(\frac
{{\nu}^{\prime}}{4}+\frac {1}{2r}\right)\left({\nu}^{\prime}-{\lambda}^{\prime}\right)\right\rbrace\right]\frac
{f_{T}}{2}-\frac {f}{4}=4\pi p_{t},\nonumber \\ \\\label{3.8}
& \qquad\hspace{-1cm} \frac{\cot\theta}{2r^{2}}T^{\prime}f_{TT}=0,
\end{eqnarray}
 where the Eq.~(\ref{3.8}) is used in the Eqs.~(\ref{3.5})-(\ref{3.7}) and hence we have the following linear form of $f(T)$ as
\begin{equation}\label{3.9}
f(T)=AT+B,
\end{equation}
where $A$ and $B$ are integration constants.

\section{Class 1 condition of the metric and general solutions}\label{sec4}
To solve the Eqs.~(\ref{3.5})-(\ref{3.7}) following \citet{Lake2003} we are choosing a specific form of the metric potential as
\begin{equation}
e^{\nu}=P(1+Er^{2})^{n},\label{4.4}
\end{equation}
where $P$ and $E$ are constants and $n < 2$.

To obtain a physically acceptable solution of the model, the metric potential $e^{\nu}$ must be positive and free from singularity with $\nu'(r)=0$ and also monotonically increasing function with $r$. From Eq.~(\ref{4.4}) we can observe that our metric function $e^{\nu}$  satisfies all the above conditions which indicates that the choice of the metric potential is physically acceptable.

Now for the spacetime of embedding class 1, Eq. (\ref{3.1}) must satisfy the Karmarkar condition~\citep{karmarkar1948}
\begin{eqnarray}
R_{1414}=\frac{R_{1212}R_{3434}+ R_{1224}R_{1334}}{R_{2323}}.\label{4.1}
\end{eqnarray}

Using the condition expressed in Eq. (\ref{4.1}) and also Eq. (\ref{3.1}), we have the following equation
\begin{eqnarray}
\frac{\lambda'\nu'}{1-e^{\lambda}}=-2(\nu''+\nu'^{2})+\nu'^{2}+\lambda'\nu', \label{4.2}
\end{eqnarray}
where $e^{\lambda}\neq 1$.

Solving the differential form given by Eq. (\ref{4.2}) and using Eq. (\ref{4.4}), we obtain
\begin{eqnarray}
e^{\lambda}=[1+F Er^2 (1+Er^2)^{(n-2)}],\label{4.5}
\end{eqnarray}
where $F=-4{n^2}PEC$ and $C$ is an integration constant.

\begin{figure}[t]
\centering
    \includegraphics[width=6cm]{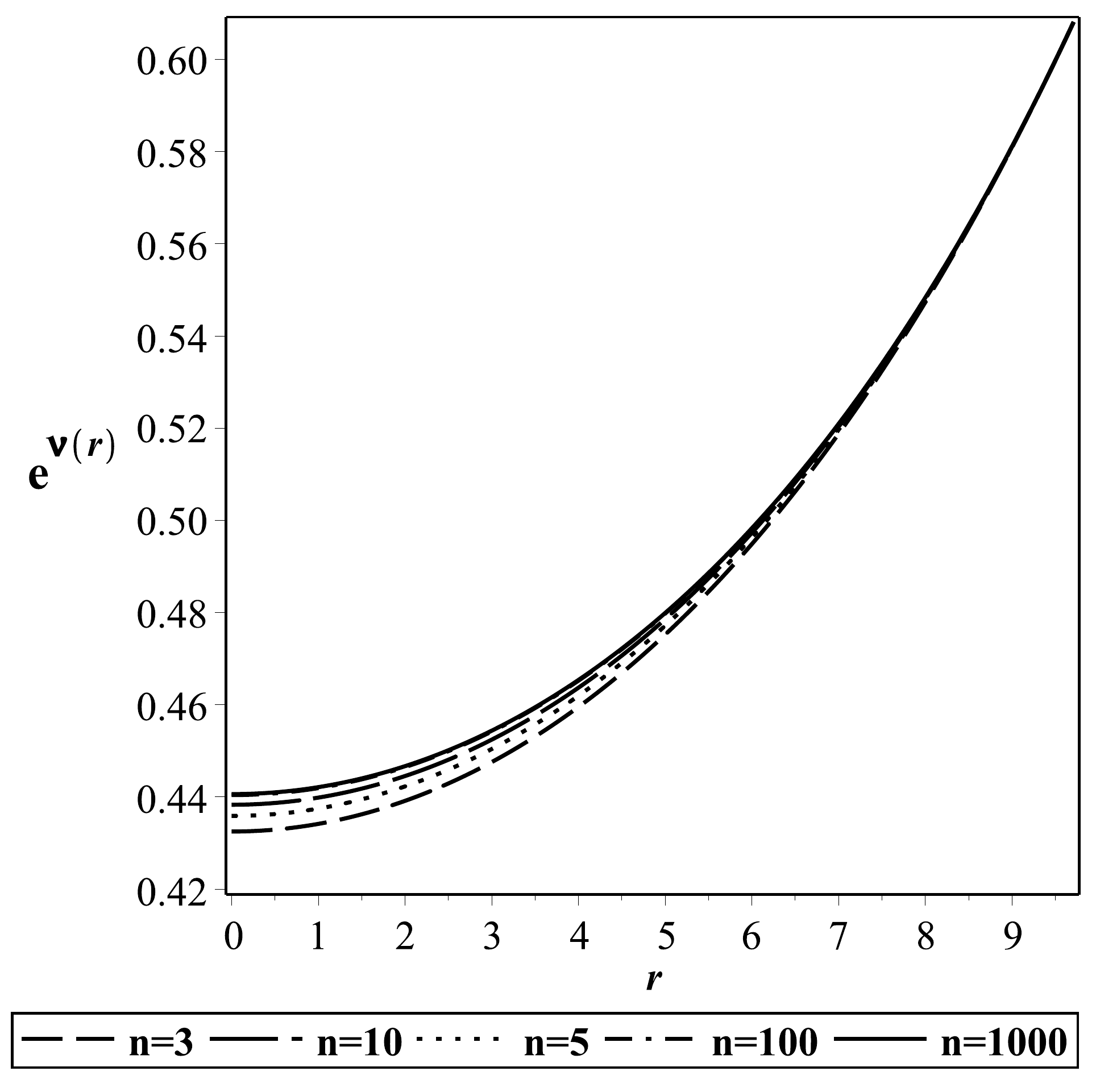}
    \includegraphics[width=6cm]{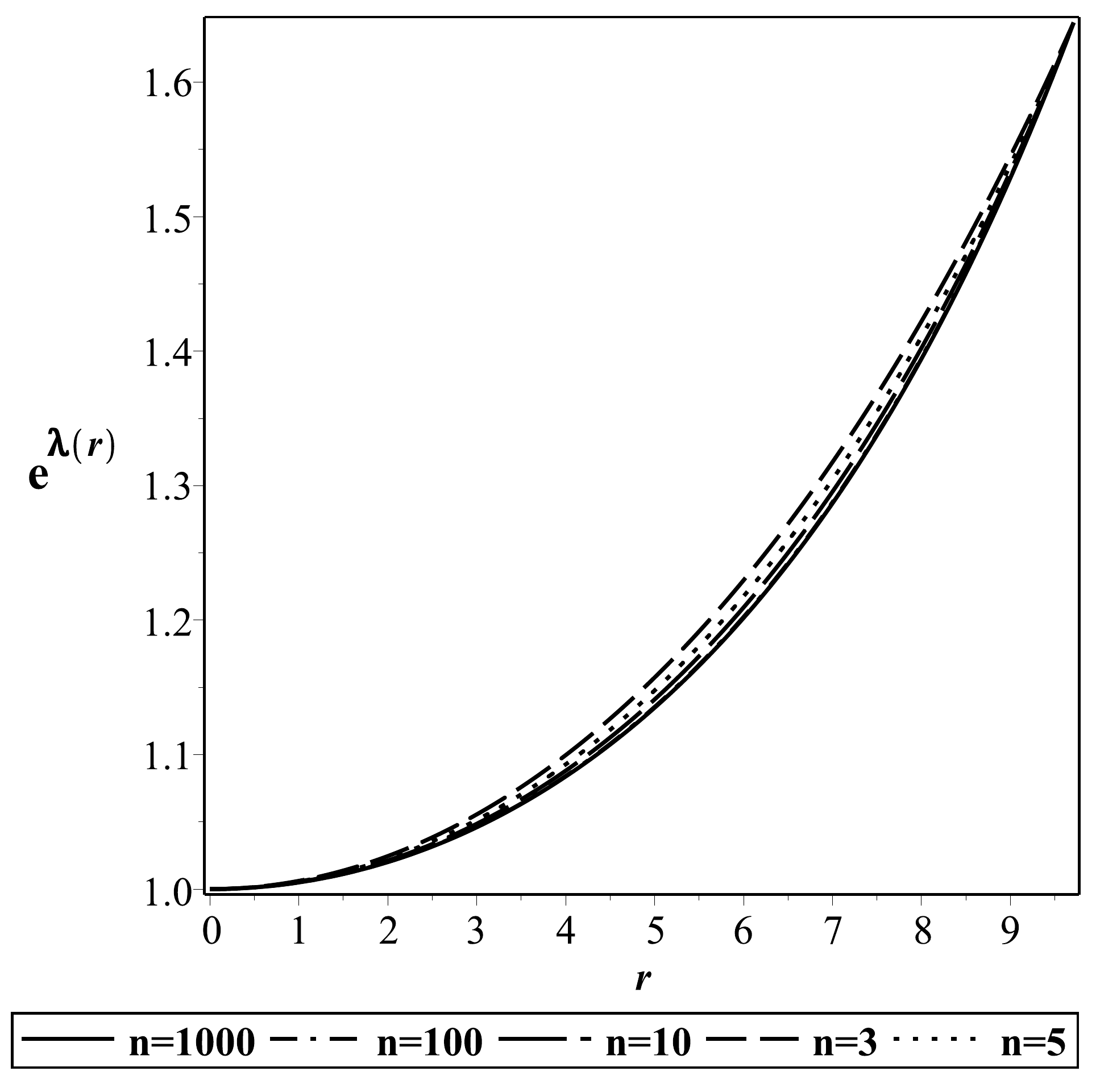}
\caption{Variation of the metric functions $e^{\nu}$ (left panel) and $e^{\lambda}$ (right panel) with the radial coordinate $r/R$ for $LMC\,X-4$.} \label{Fig1}
\end{figure}

Now solving Eqs. (\ref{3.5})-(\ref{3.7}), (\ref{4.4}) and (\ref{4.5}) we have the expressions of the energy density ($\rho$), radial pressure ($p_r$) and tangential pressure ($p_t$) given as
\begin{eqnarray}\label{4.6}
&\qquad\hspace{-1cm} \rho={\frac {A \left[ 2 \left( E{r}^{2}+1 \right) ^{n+1}\left\lbrace \frac{3}{2}+ \left( n-\frac{1}{2} \right) {r}^{2}E \right\rbrace
+FE{r}^{2} \left( E{r}^{2}+1 \right)^{2n} \right] EF}{ 8\pi \left[
 \left( E{r}^{2}+1 \right) ^{2}+FE{r}^{2} \left( E{r}^{2}+1 \right)^{
n} \right] ^{2} }}, \\ \label{4.7}
&\qquad\hspace{-1cm} p_{{r}}={\frac {AE \left[ 2E{r}^{2}n-F \left(E{r}^{2}+1
 \right) ^{n}+2n \right] }{8\pi \left[\left( E{r}^{2}+1 \right)
^{2}+FE{r}^{2} \left( E{r}^{2}+1 \right)^{n} \right] }},\\ \label{4.8}
&\qquad\hspace{-1cm} p_{{t}}={\frac {A \left[ F \left( E{r}^{2}-1 \right)  \left( E{r}
^{2}+1 \right)^{n}+n \left( E{r}^{2}+1 \right)  \left( E{r}^{2}n+2
 \right)  \right] E \left( E{r}^{2}+1 \right) }{8\pi \left[  \left(
E{r}^{2}+1 \right) ^{2}+FE{r}^{2} \left( E{r}^{2}+1 \right)^{n}
 \right]^{2}}}.  
 \end{eqnarray}

\begin{figure*}[!htp]
\centering
    \includegraphics[width=6cm]{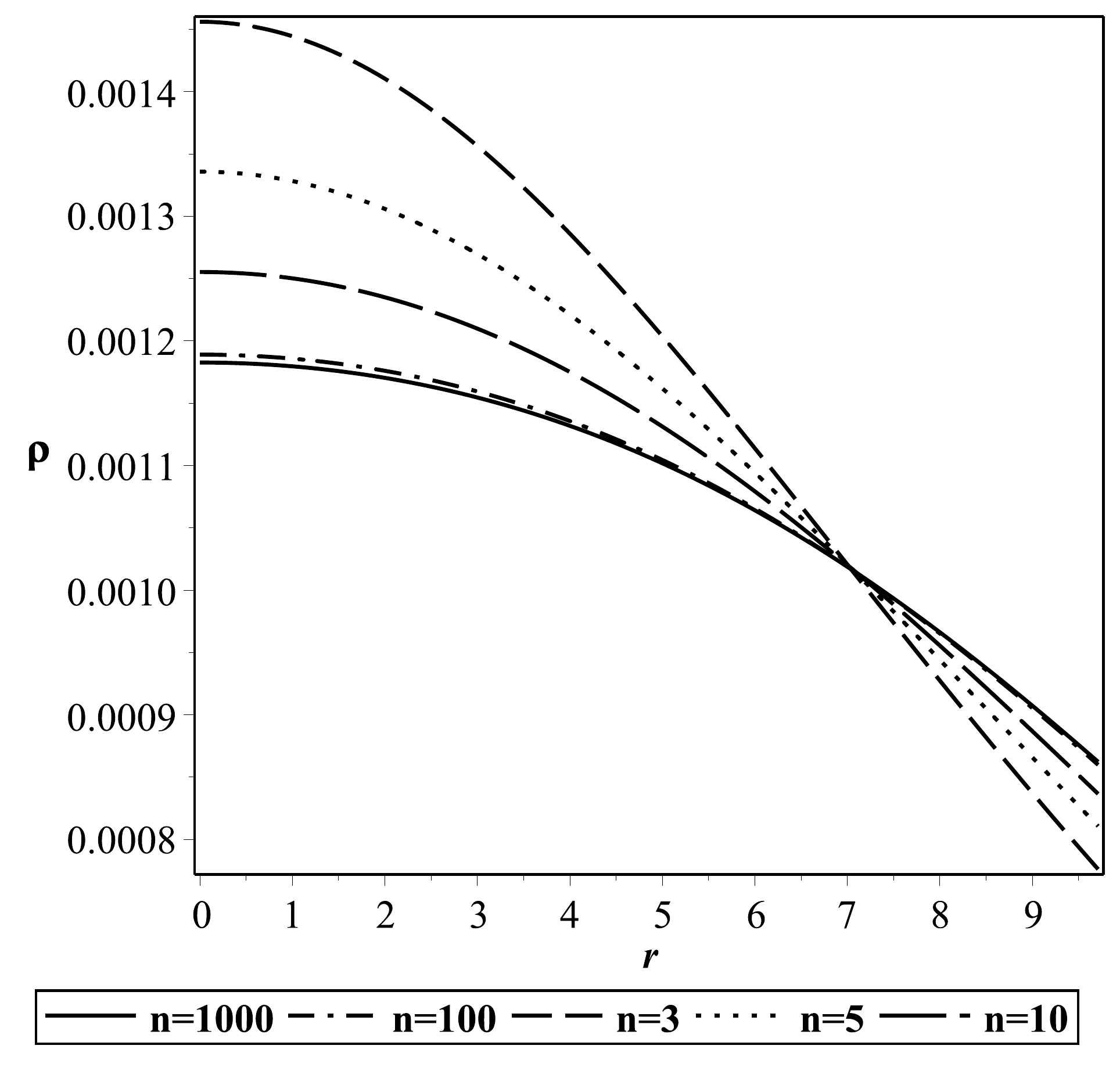}
    \caption{Variation of the matter density with the radial coordinate $r/R$ for $LMC\,X-4$.} \label{Fig2}
\end{figure*}

\begin{figure*}[t]
\centering
    \includegraphics[width=6cm]{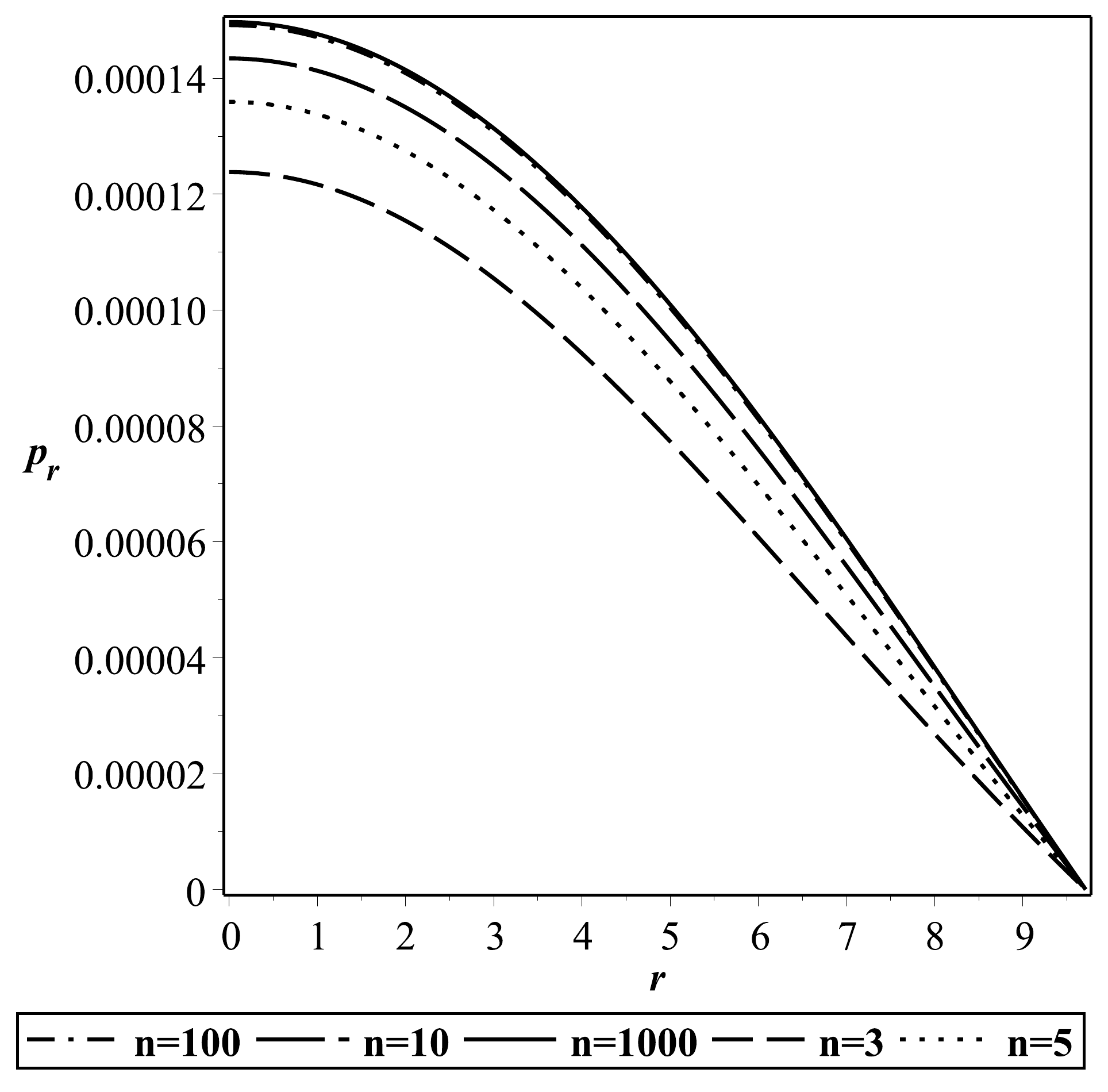}
    \includegraphics[width=6cm]{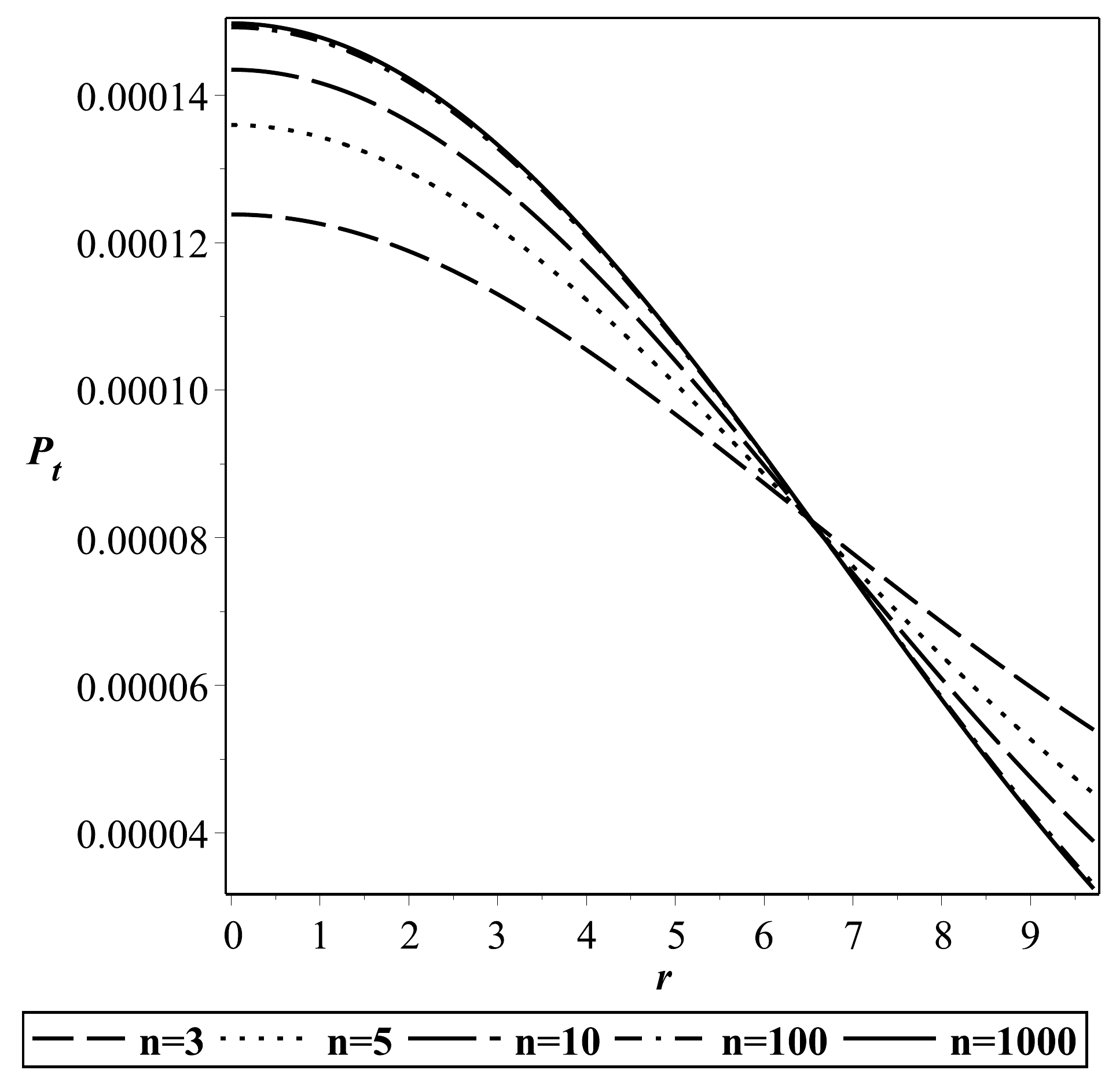}
\caption{Variation of the radial pressure $p_r$ (left panel) and tangential pressure $p_t$ (right panel) with the radial coordinate $r/R$ for $LMC\,X-4$.} \label{Fig3}
\end{figure*}

The variation of the physical parameters, such as the density, radial pressure and tangential pressure with respect to the radial coordinate $r$ inside the stellar configuration for different values of n are featured in Figs.~\ref{Fig2}~and~\ref{Fig3}. The density (Fig.~\ref{Fig2}) and the pressures (Fig.~\ref{Fig3}) are maximum at the centre and decreases monotonically inside the stellar system to reach the minimum value at the surface which confirms physical validity of the solutions. 

Using Eqs. (\ref{4.7}) and (\ref{4.8}), the anisotropy ($\Delta$) of the stellar system can be given as
\begin{eqnarray}\label{4.9}
& \qquad \Delta={\frac { {r}^{2}A{E}^{2} \left[ -{r}^{2} \left( n-2 \right) E+F g ^{n}-n+2 \right]  \left[ -E{r}^{2}n+F
g ^{n}-n \right] }{8\pi \left[  g ^{2}+FE{r}^{2} g ^{n}
 \right] ^{2}}}, 
\end{eqnarray}
where $g=\left(E{r}^{2}+1\right)$. 

\begin{figure*}[!htp]
\centering
    \includegraphics[width=6cm]{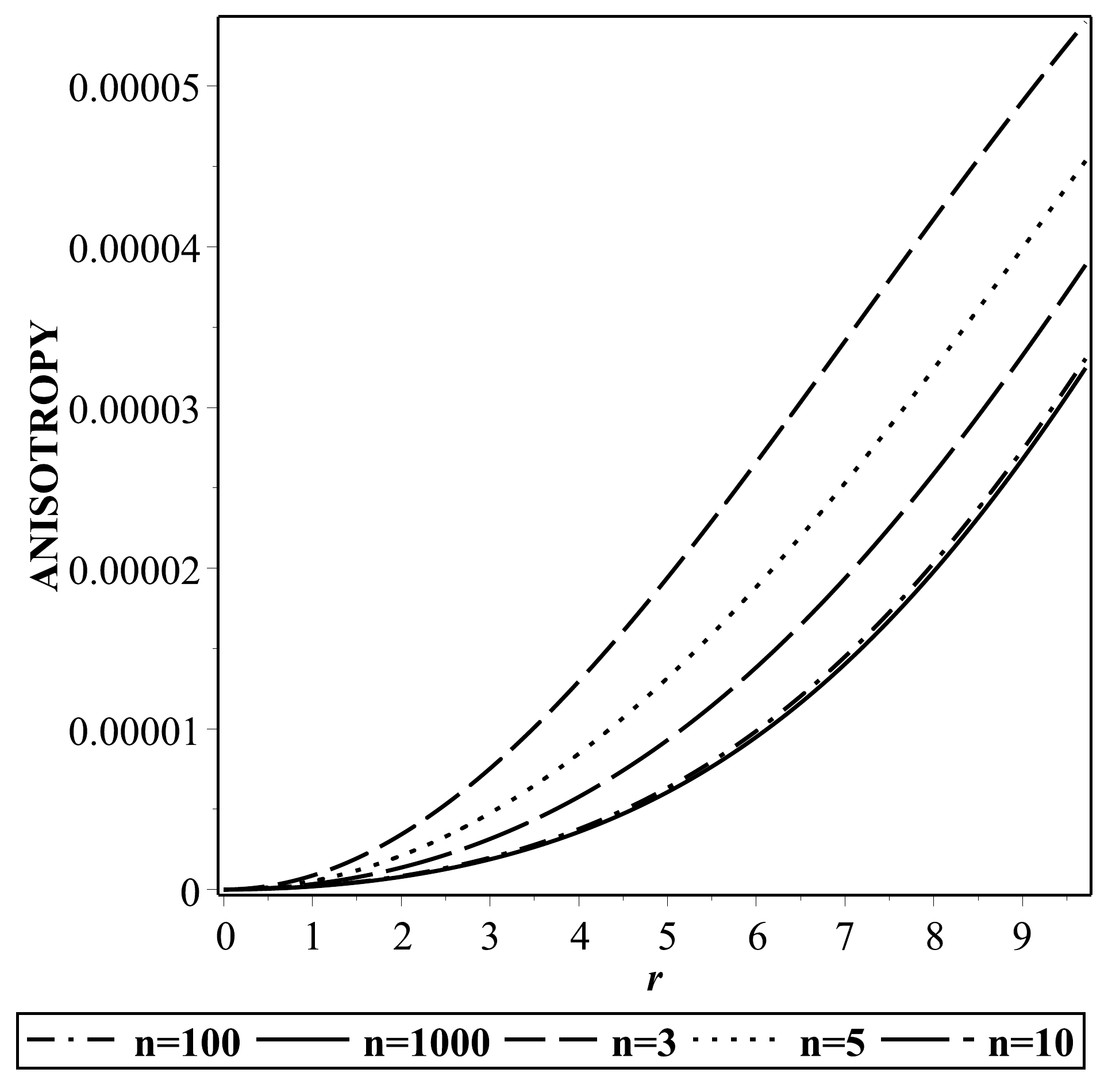}
    \caption{Variation of the anisotropy factor with the radial coordinate $r/R$ for $LMC\,X-4$.} \label{Fig4}
\end{figure*}

We find from Fig.~\ref{Fig4} that the anisotropy is minimal, i.e., zero at the centre and maximal at the surface as predicted by~\citet{Deb2016}.

From our model the radial (${\omega}_r$) and tangential (${\omega}_t$) equation of state (EOS) parameters are given as
\begin{eqnarray}\label{4.10}
&\qquad\hspace{-1cm} \omega_{{r}}=\frac{p_r}{\rho}={\frac { \left[ 2E{r}^{2}n-F \left( E{r}^{2}+1 \right)^{n}+2n \right]  \left[  \left( E{r}^{2}+1 \right)^{2}+FE{r}^{2} \left( E{r}^{2}+1 \right)^{n} \right] }{ \left[ 2 \left( E{r}^{2}+1 \right)^{n+1} \left\lbrace \frac{3}{2}+ \left( -\frac{1}{2}+n \right) {r}^{2}E \right\rbrace +FE{r}^{2} \left( E{r}^{2}+1 \right)^{2n} \right] F}}, \\ \label{4.11}
 &\qquad\hspace{-1cm} \omega_{{t}}=\frac{p_t}{\rho}={\frac { \left[ F \left( E{r}^{2}-1 \right)  \left( E{r}^{2}+1 \right)^{n}+n \left( E{r}^{2}+1 \right)  \left( E{r}^{2}n+2 \right)  \right]  \left( E{r}^{2}+1 \right) }{ \left[ 2 \left( E{r}^{2}+1 \right) ^{n+1} \left\lbrace \frac{3}{2}+ \left( -\frac{1}{2}+n \right) {r}^{2}E \right\rbrace +FE{r}^{2} \left( E{r}^{2}+1 \right)^{2n} \right] F}}. 
\end{eqnarray}

\begin{figure*}[!htp]
\centering
    \includegraphics[width=6cm]{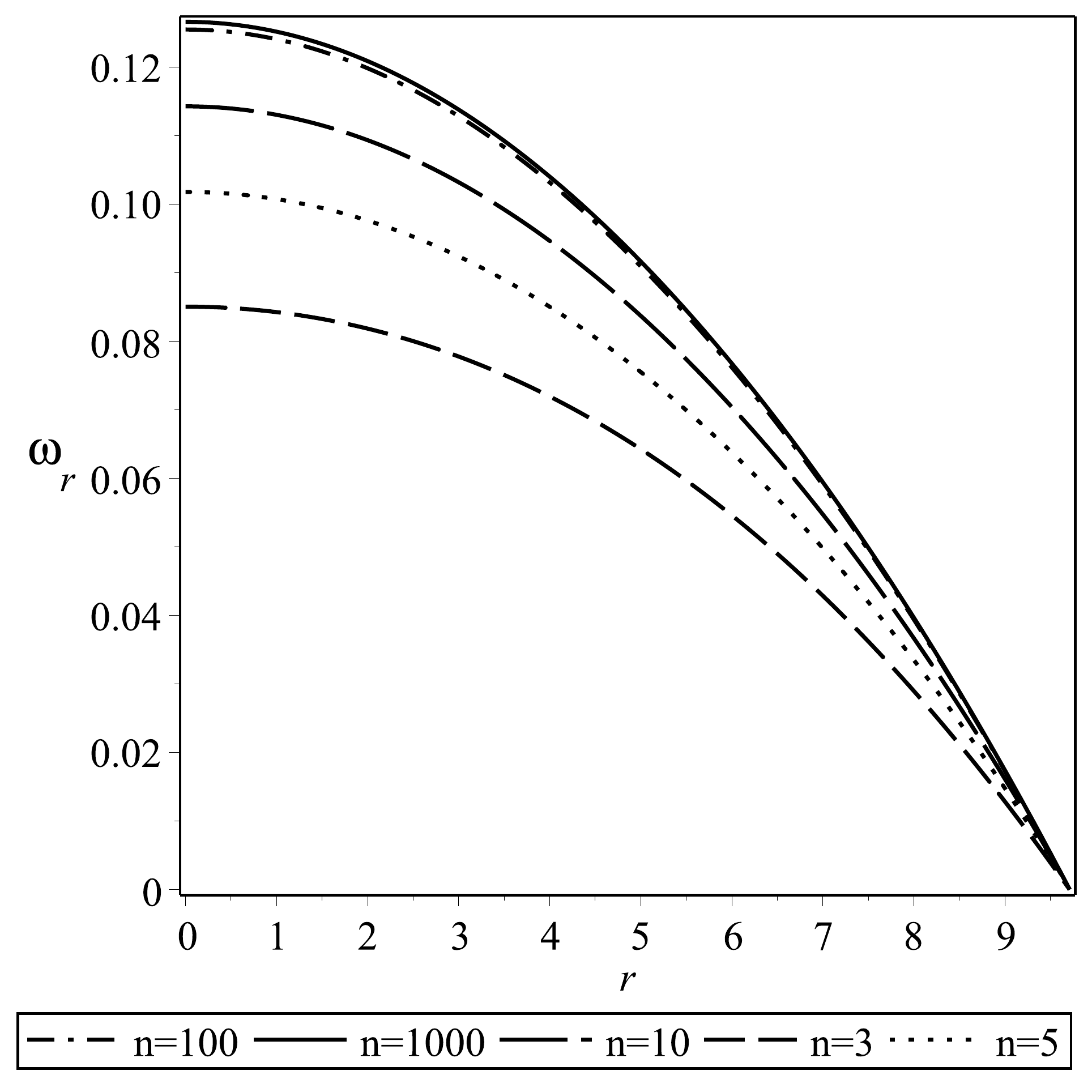}
    \includegraphics[width=6cm]{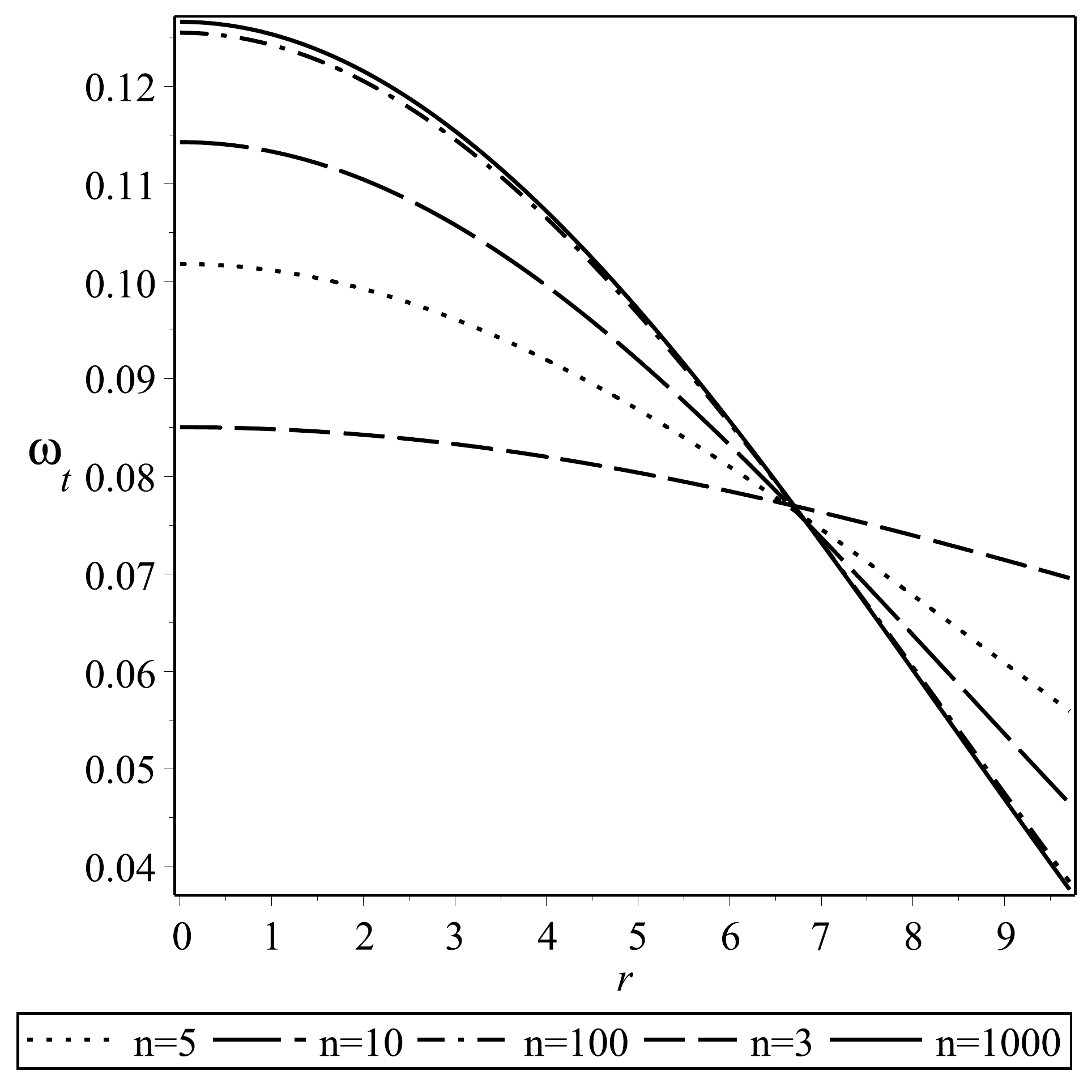}
\caption{Variation of $\omega_r$ and $\omega_t$ with the radial coordinate $r$ for $LMC\,X-4$.} \label{Fig4.1}
\end{figure*}

The variation of radial EOS ($\omega_r$) and tangential EOS ($\omega_t$) with respect to the radial coordinate $r$ are presented in Fig.~\ref{Fig4.1}. 

\section{Boundary Conditions}\label{sec5}
The exterior spacetime of the star can be described by the Schwarzschild metric, given by
\begin{eqnarray}\label{5.1}
&\qquad\hspace{-0.5cm} ds^{2} =\left(1-\frac{2M}{r} \right) dt^{2} -\frac{dr^{2}}{\left(1-\frac{2M}{r}\right)}-r^{2} (d\theta^{2}+\sin ^{2} \theta d\phi ^{2} ).\nonumber \\
\end{eqnarray}

Now to calculate the values of the constants, viz., $E$, $P$, $C$, $F$ and $B$ we match our interior spacetime with the exterior one as given in \ref{5.1} to get the following equalities
\begin{eqnarray}\label{5.2}
&\qquad P \left( E{R}^{2}+1 \right) ^{n}=1-{\frac {2M}{R}},\\\label{5.3}
&\qquad 1+FE{R}^{2} \left( E{R}^{2}+1 \right) ^{n-2}= \left( 1-{\frac {2M}{R}} \right)^{-1},\\\label{5.4}
&\qquad P \left( E{R}^{2}+1 \right)^{n-1}nE={\frac {M}{{R}^{3}}}.
\end{eqnarray}

Also for a physically acceptable stellar model the radial pressure must be zero at the boundary, i.e., at $r=R$ we have
\begin{equation}
{p_r}(R)=0. \label{5.6}
\end{equation}

Using Eqs. (\ref{5.2})-(\ref{5.6}) we have following expressions of the constants $E$, $P$, $C$, $F$ and $B$ given as
\begin{eqnarray}\label{5.7}
&\qquad E=-{\frac {M}{{R}^{2} \left( 2\,Mn-nR+M \right) }},\\\label{5.8}
&\qquad P= \left( 1-{\frac {2M}{R}} \right)  \left[ {\frac {n \left( 2\,M-R
 \right) }{2\,Mn-nR+M}} \right] ^{-n},\\\label{5.9}
&\qquad F={\frac {2{n}^{2} \left( 2\,M-R \right) }{2\,Mn-nR+M} \left[ {
\frac {n \left( 2\,M-R \right) }{2\,Mn-nR+M}} \right] ^{-n}},\\\label{5.10}
&\qquad C=-{\frac {{R}^{3}}{2M}},\\\label{5.11}
&\qquad B=0.
\end{eqnarray}

\section{Physical Properties of the model}\label{sec6}
The physical parameters of the anisotropic stellar model are studied in the following subsections.

\subsection{Energy conditions}\label{sub6.1}
The energy conditions, viz., Weak Energy Condition (WEC), Null Energy Condition (NEC), Strong Energy
Condition (SEC) and Dominant Energy Condition (DEC) for an anisotropic fluid sphere are hold
if and only if the following inequalities are satisfied simultaneously at every points in the interior of the fluid sphere:
\begin{eqnarray}
NEC: \rho+p_r\geq 0,~\rho+p_t\geq 0, \label{6.1.1}\\
WEC: \rho+p_r\geq 0,~\rho\geq 0,~\rho+p_t\geq 0, \label{6.1.2}\\
SEC:  \rho+p_r\geq 0,~\rho+p_r+2p_t\geq 0,\label{6.1.3}\\
DEC: \rho\geq0,~ \rho\pm {p_r}\geq0,~~ \rho\pm {p_t}\geq0. \label{6.1.3}
\end{eqnarray}

\begin{figure*}[t]
\centering
    \subfloat{\includegraphics[width=4.5cm]{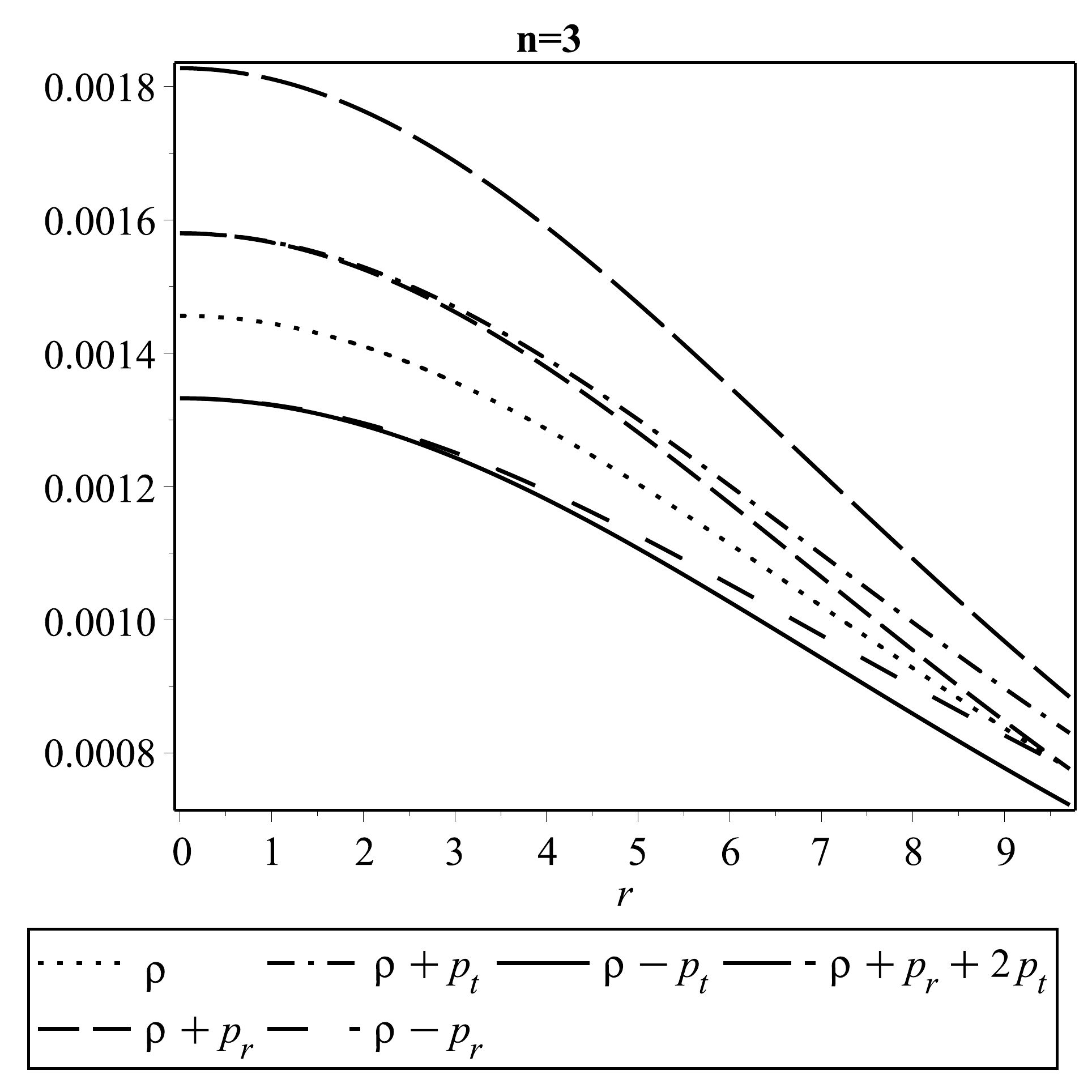}}
    \subfloat{\includegraphics[width=4.5cm]{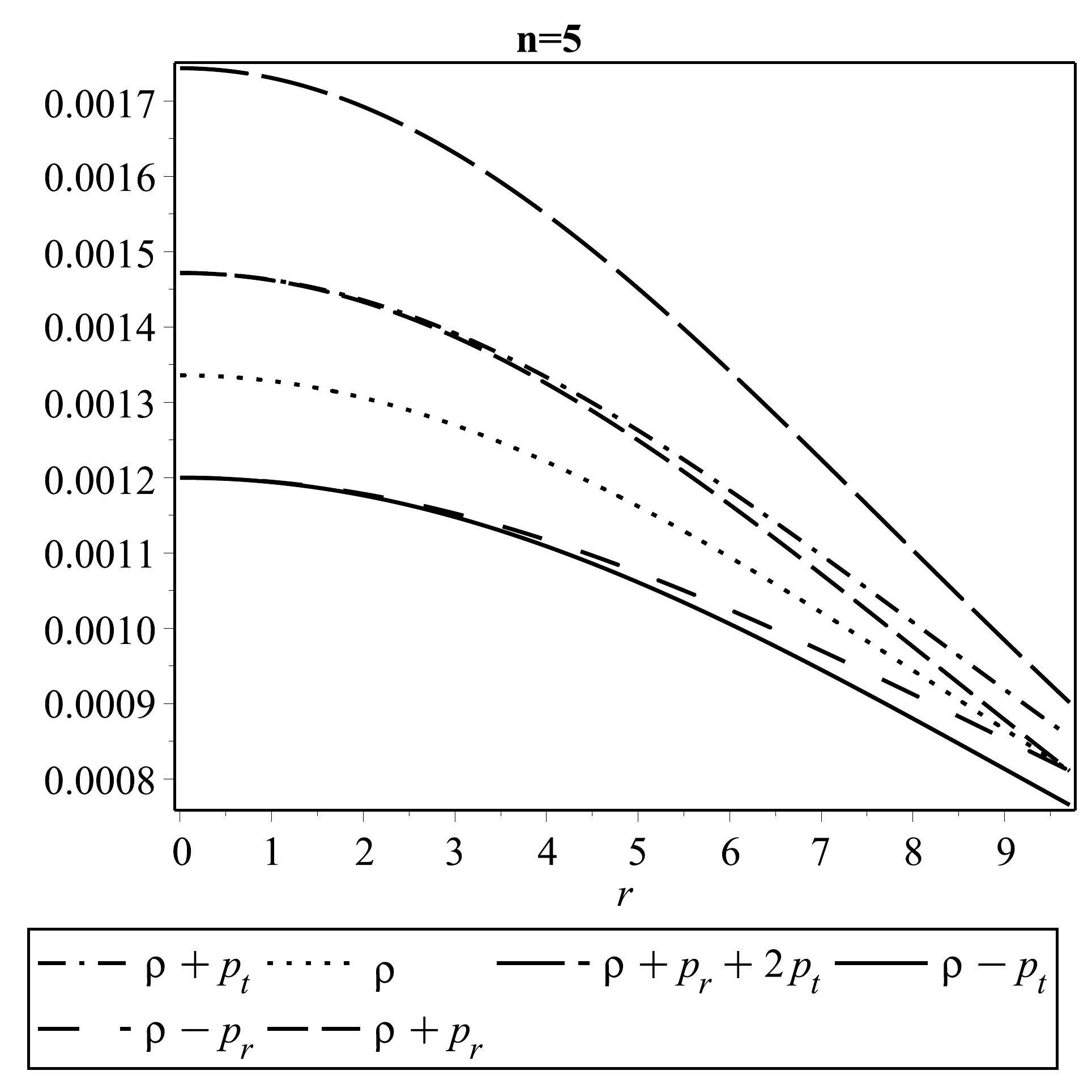}}
    \subfloat{\includegraphics[width=4.5cm]{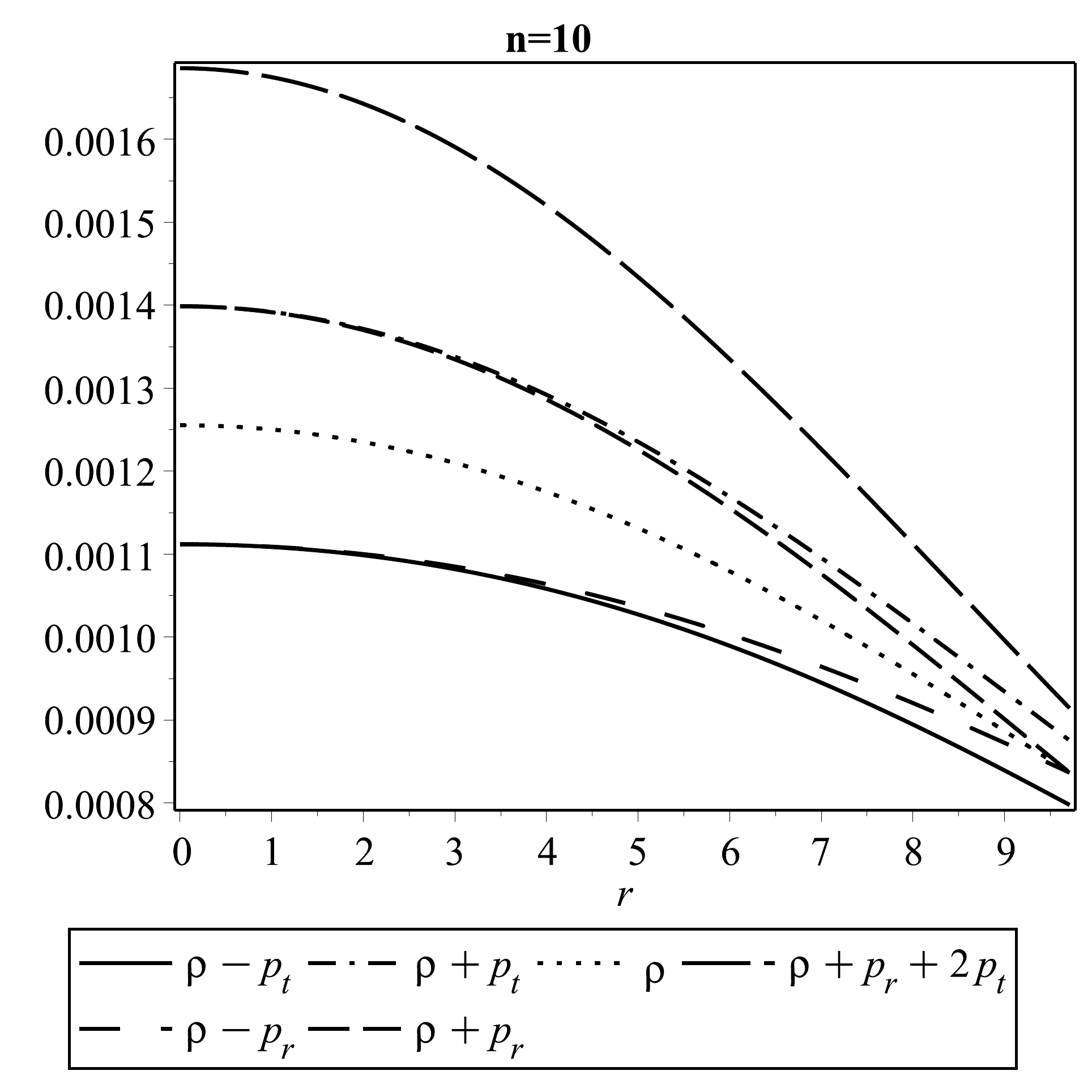}}
    \
    \subfloat{\includegraphics[width=5cm]{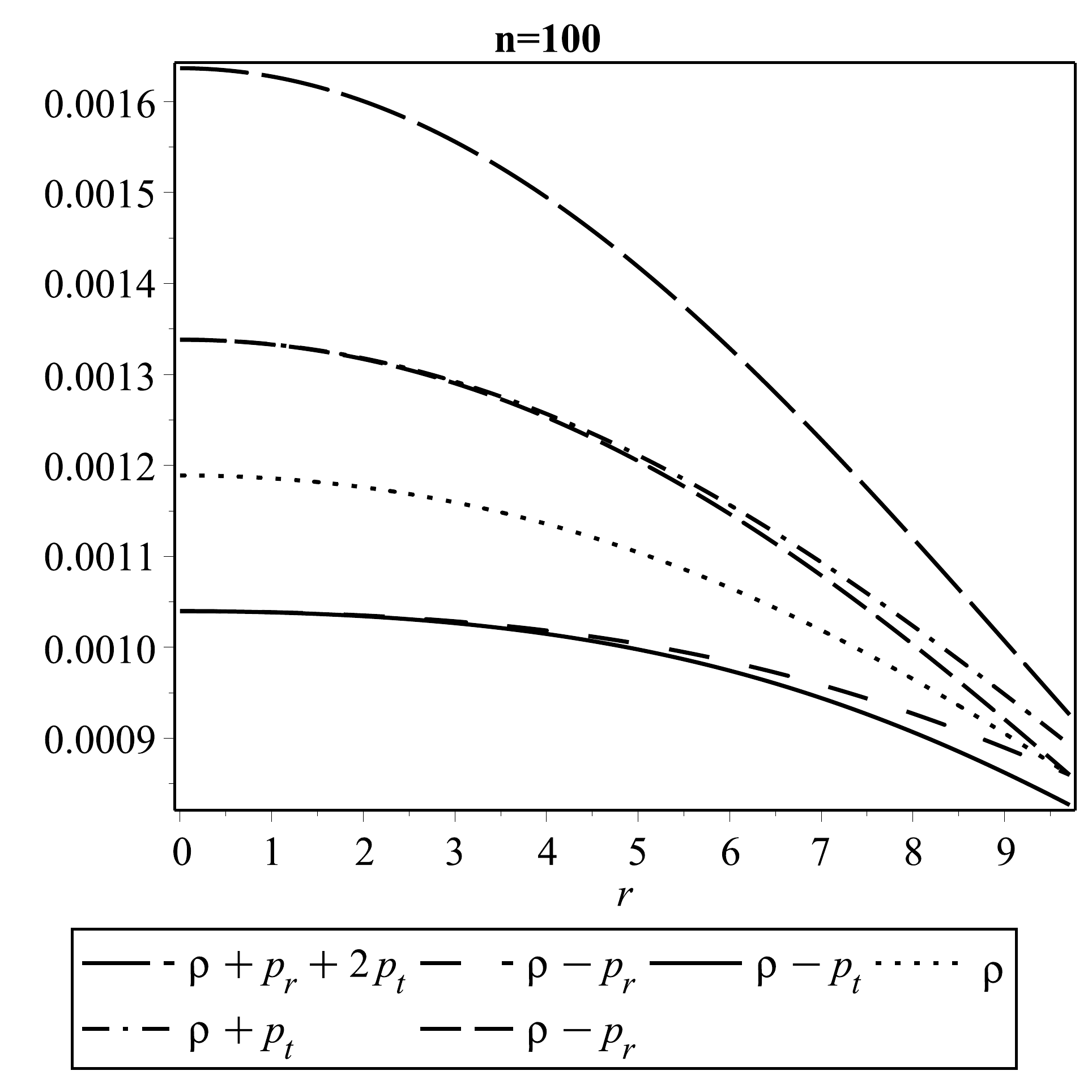}}
    \subfloat{\includegraphics[width=5cm]{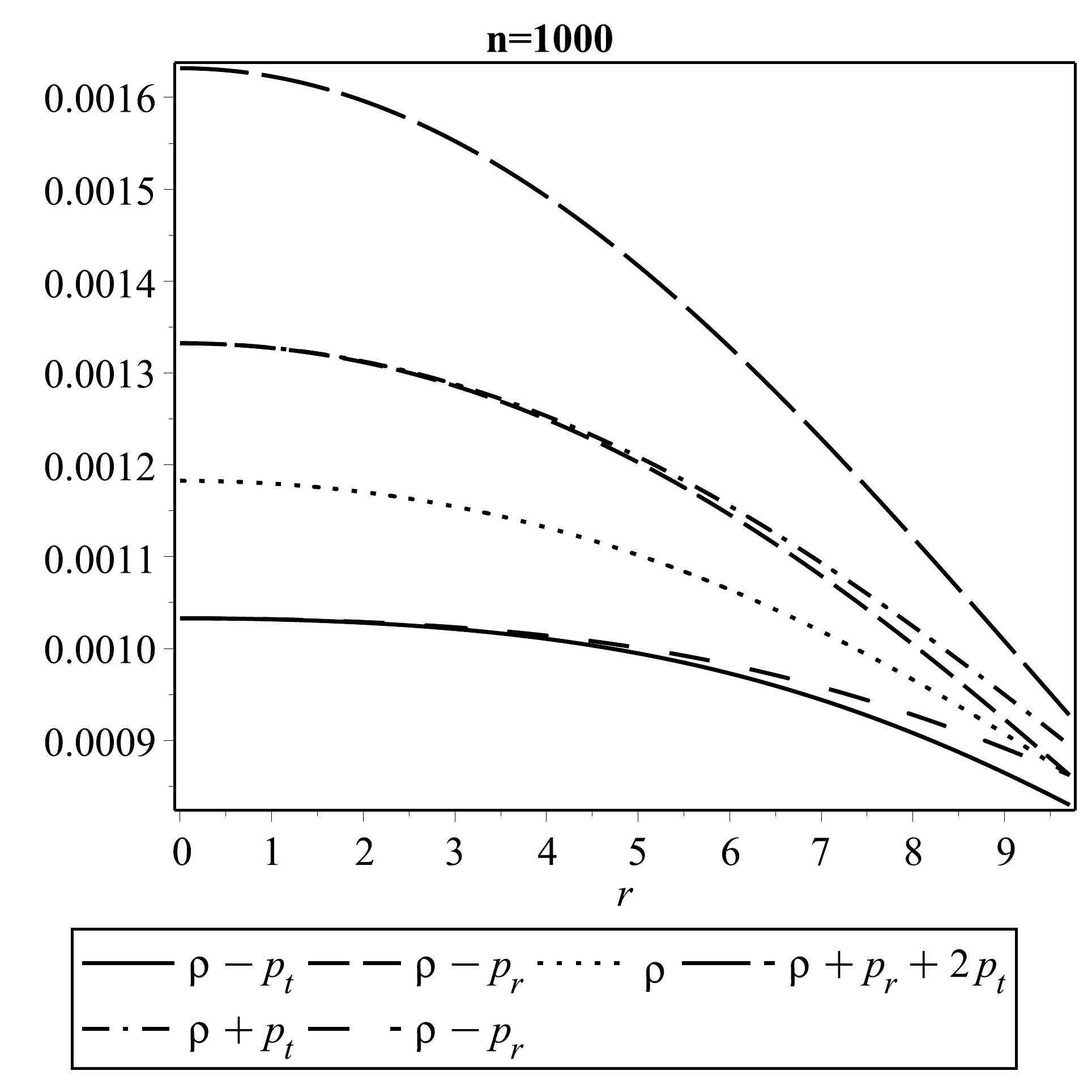}}
    \caption{Variation of energy conditions with the radial coordinate $r/R$ for $LMC\,X-4$.} \label{Fig5}
\end{figure*}

Fig.~\ref{Fig5} features that our model is consistent with all the energy conditions. 

\subsection{Stability of the model}\label{sub6.2}
Now our task is to examine the stability of our proposed model. In the following sections
we are going to discuss stability of the stellar model.

\subsubsection{Generalized TOV equation}\label{subsub6.2.1}
~\citet{Tolman1939}, and later on~\citet{Oppenheimer1939}
suggested that a physically acceptable model must be stable under
the three forces, viz., the gravitational force $(F_g)$, the hydrostatic
force $(F_h)$ and the anisotropic force $(F_a)$ in such a way that the
sum of these three forces becomes zero in equilibrium, i.e.
\begin{equation}
F_g+F_h+F_a=0. \label{6.2.1.1}
\end{equation}

Now we can construct the generalised Tolman-Oppenheimer-Volkoff (TOV) equation~\citep{Varela2010,Ponce1993} for our system as
\begin{equation}
-\frac{M_G(r)(\rho+p_r)}{r^2}e^{\frac{\lambda-\nu}{2}}-\frac{dp_r}{dr}+\frac{2}{r}(p_t-p_r)=0, \label{6.2.1.2}
\end{equation}
where the gravitational mass  within the radius $r$ is represented by $M_G(r )$. 

Using the Tolman-Whittaker formula~\citep{Devitt1989} and also the Einstein field equations, $M_G(r)$ can be written in the following form
\begin{equation}
M_G(r)=\frac{1}{2}{{r}^{2}}e^{\frac{\nu-\lambda}{2}}\nu'. \label{6.2.1.3}
\end{equation}

 Putting the value of Eq. (\ref{6.2.1.3}) in Eq. (\ref{6.2.1.2}) we have
\begin{equation}
-\frac{\nu'}{2}(\rho+p_r)-\frac{dp_r}{dr}+\frac{2}{r}(p_t-p_r)=0. \label{6.2.1.4}
\end{equation}

\begin{figure*}[!htp]
\centering
    \subfloat{\includegraphics[width=4.5cm]{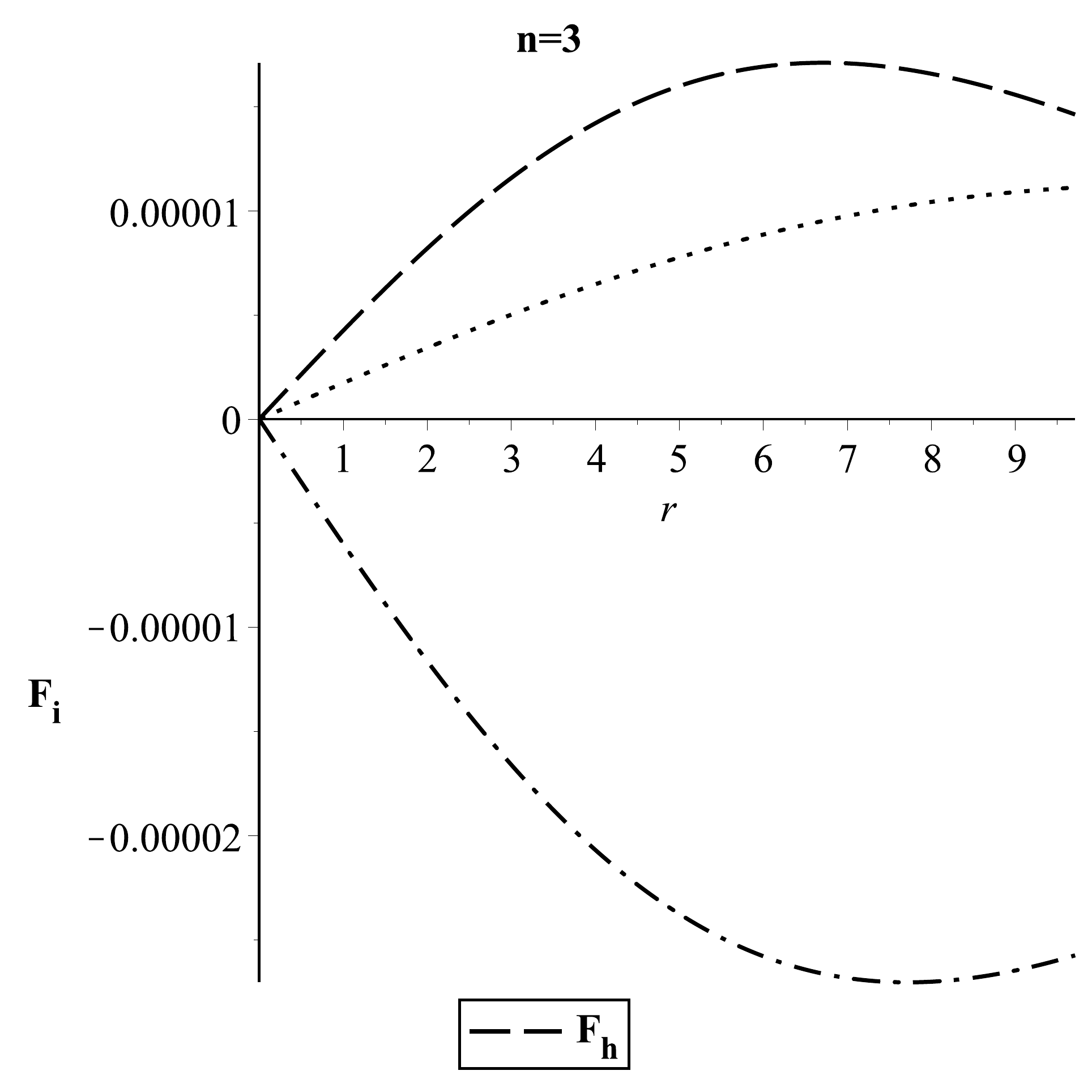}}
    \subfloat{\includegraphics[width=4.5cm]{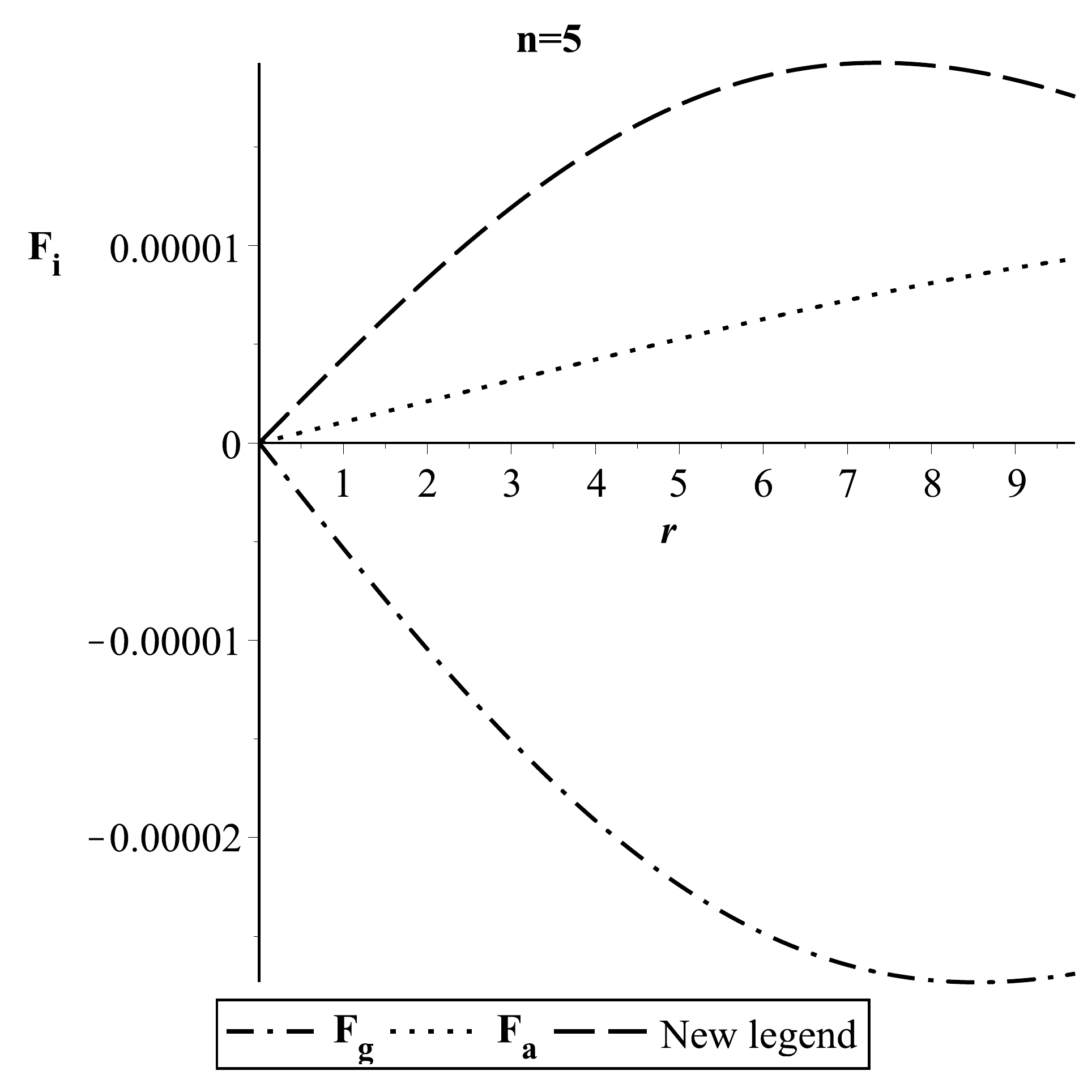}}
    \subfloat{\includegraphics[width=4.5cm]{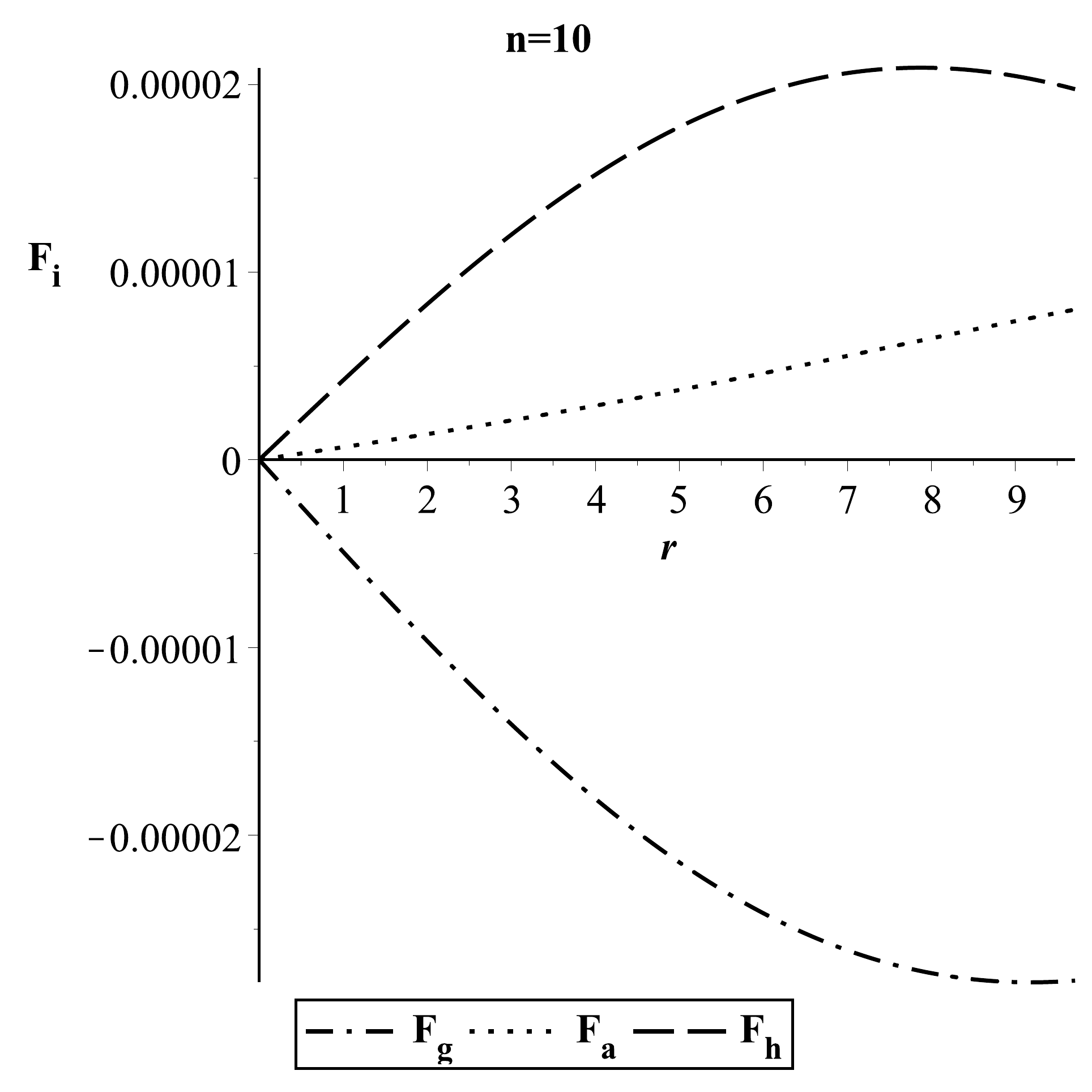}}
    \
    \subfloat{\includegraphics[width=4.5cm]{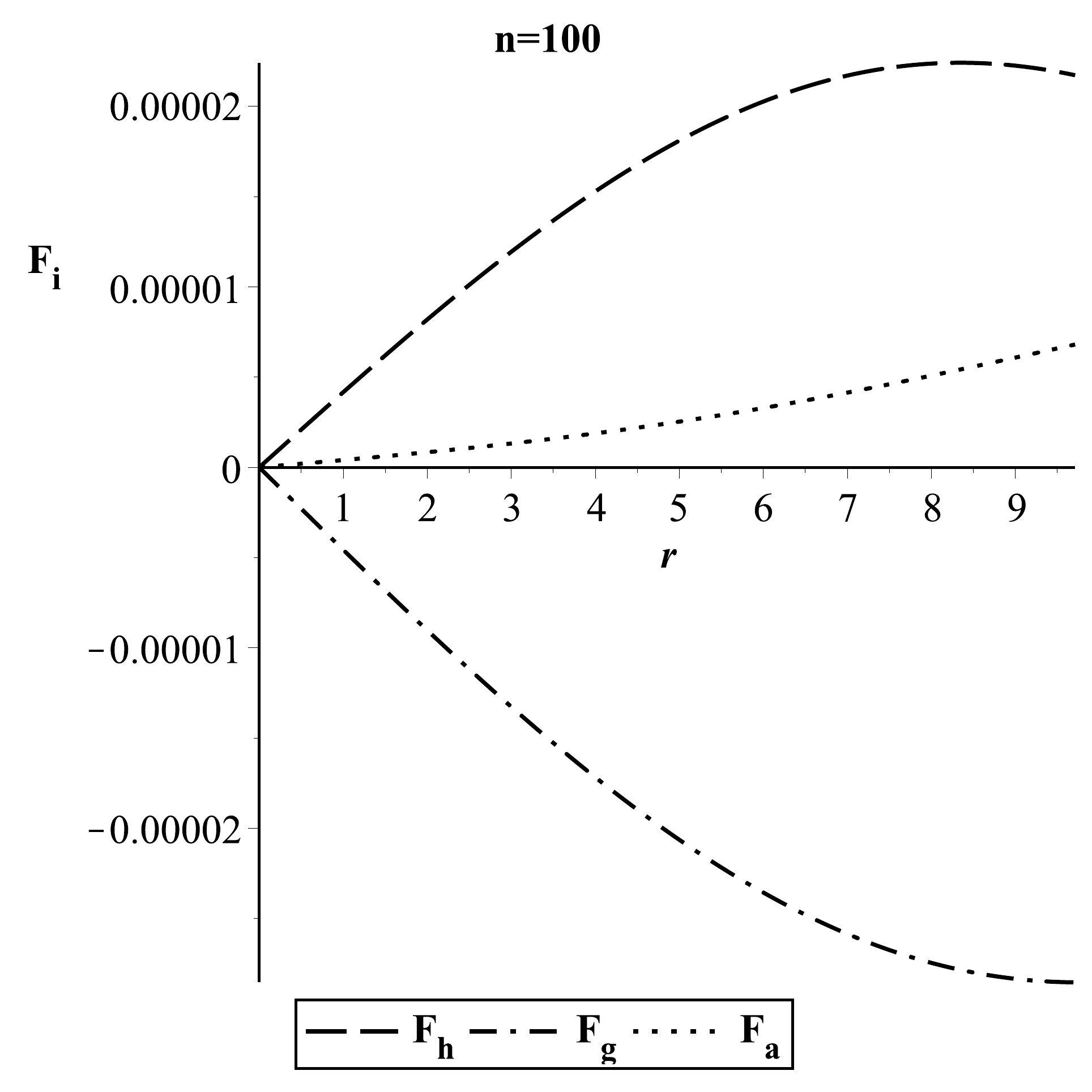}}
    \subfloat{\includegraphics[width=4.5cm]{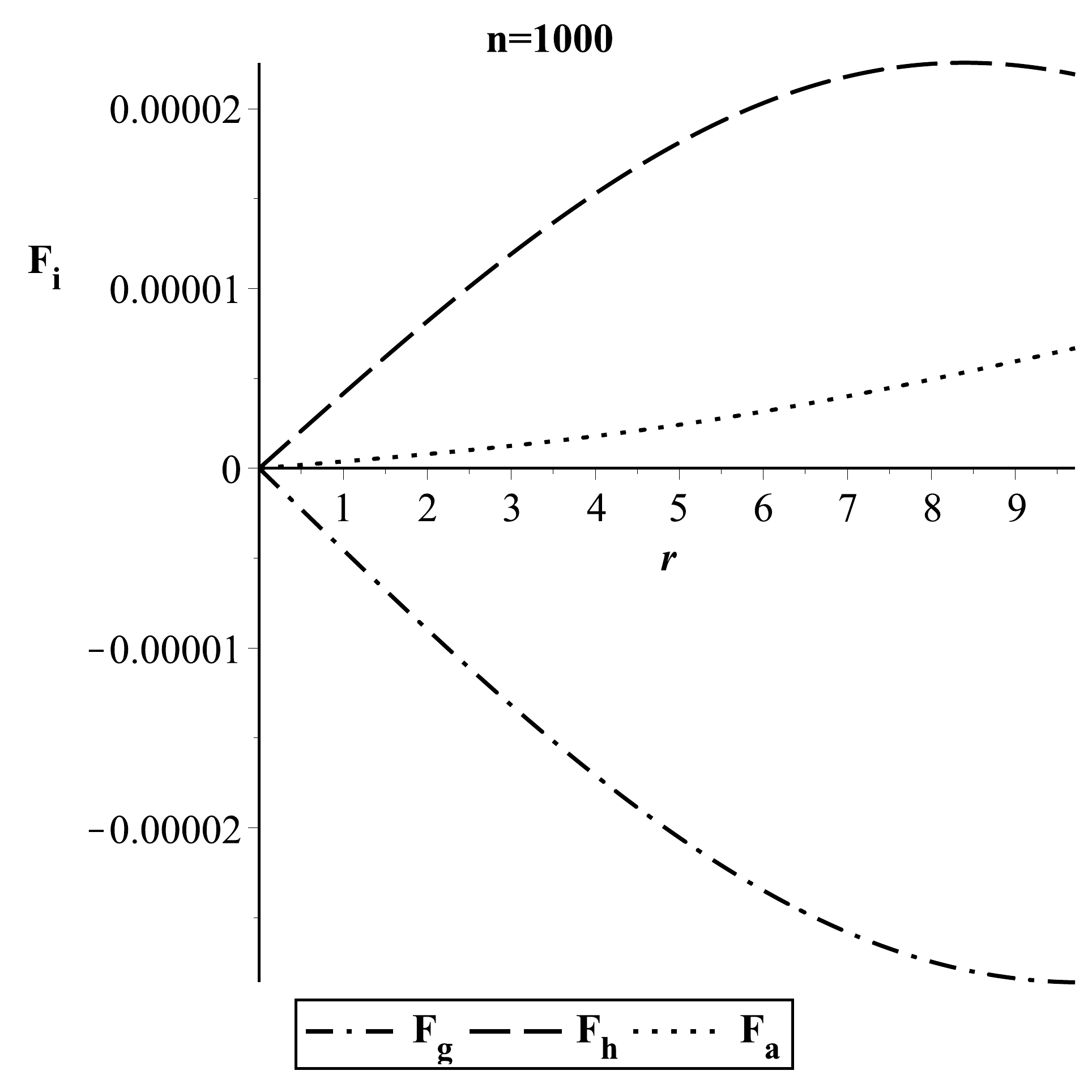}}
   \caption{Variation of the forces with the radial coordinate $r/R$ for $LMC\,X-4$.} \label{Fig6}
\end{figure*}

Now from Eqs. (\ref{6.2.1.1}) and (\ref{6.2.1.4}), we can write the forces as
\begin{eqnarray}\label{6.2.1.5}
&\qquad\hspace{-1.4cm} {F_g}=-\frac{\nu'}{2}(\rho+p_r)=-{\frac {Ar{E}^{2} \left[ 2 \left\lbrace \frac{1}{2}+ \left( n-\frac{1}{2} \right) {r}^{2}E \right\rbrace F g ^{n}+n g ^{2} \right] n}{4\pi \left(  g ^{2}+FE{r}
^{2} g ^{n} \right) ^{2} }},  \\\label{6.2.1.6}
&\qquad\hspace{-1.4cm} {F_h}=-\frac{dp_r}{dr}={\frac {Ar{E}^{2} \left[ -\frac{1}{2}{F}^{2} g
^{2\,n}+ \left\lbrace {r}^{2} \left( {n}^{2}+\frac{n}{2}-1 \right) E+\frac{3}{2}n-1
 \right\rbrace F g ^{n}+n g^{2} \right] }{2\pi \left(  g ^{2}+FE{r}^{2}
 g ^{n} \right) ^{2}}},\nonumber \\ \\\label{6.2.1.7}
&\qquad\hspace{-1.4cm} {F_a}=\frac{2}{r}(p_t-p_r)={\frac {Ar{E}^{2} \left[ {F}^{2} g^{2n
}+ \left\lbrace -2F \left( n-1 \right)  g ^{n}+ \left( n-2 \right)  g n \right\rbrace  g  \right] }{4\pi \left[  g ^{2}+FE{r}^{2} g^{n} \right] ^{2}}},
\end{eqnarray}
where $g=\left(E{r}^{2}+1\right)$.

The variation of these three forces with the fractional radial coordinate, however after taking different values of $n$, for $LMC~X-4$ is shown in Fig.~\ref{Fig6} which demonstrates that as the equilibrium of the forces is achieved our system is stable.

\subsubsection{Herrera cracking concept}\label{subsub6.2.2}
Any stable configuration of an anisotropic fluid distribution must
satisfy the condition proposed by \citet{Herrera1992}, known as Herrera's
cracking condition. According to this condition the radial sound speed $(v_r^2)$ and the tangential sound speed
$(v_t^2)$ must lie between $0$ and $1$. Again following \citet{Herrera1992} and \citet{Abreu2007}, the provided condition for the stable region is $|v_t^2-v_r^2|\leq1$. The radial $(v_r^2)$ and tangential $(v_t^2)$ sound speeds for our system are given as
\begin{eqnarray}\label{6.2.2.1}
& \qquad\hspace{-1.5cm}{v^{2}_{{r}}}={\frac {\rm d\,{p_r}}{{\rm d}\rho}} = \Bigg[-{F}^{3}{g}^{2\,n}E{r}^{2}+2 [ -\frac{1}{2}  + \left( n+\frac{3}{2} \right)\nonumber \\
& {r}^{4} \left( n-1 \right) {E}^{2} +\frac{3}{2} \left( n-\frac{4}{3} \right) {r}^{2}E ] {F}^{2}{g}^{n} +2[\left( -\frac{1}{2}+n \right)\nonumber \\
&  \left( n+2 \right) {r}^{2}E +\frac{3}{2}n-1 ] {g}^{2}F+2\,n{g}^{4-n}\Bigg]\Bigg/\Bigg[2F\nonumber \\
& \big[ \frac{1}{2}{F}^{2}{g}^{2\,n}E{r}^{2}+F \lbrace \frac{5}{2}+{r}^{4}
 \left( {n}^{2}-\frac{3}{2}n+\frac{3}{2} \right) {E}^{2}\nonumber \\
& +\frac{3}{2}\left( n+\frac{4}{3}\right)
{r}^{2}E \rbrace {g}^{n} -{g}^{2} \left\lbrace \frac{5}{2}+ \left( -\frac{1}{2}+n \right) {r}
^{2}E \right\rbrace \left( n-2 \right)  \big]\Bigg], \\ \label{6.2.2.2}
& {v^{2}_{{t}}}={\frac {\rm d\,{p_t}}{{\rm d}\rho}}=
  \Bigg[n{g}^{3} \left( E{r}^{2}n-n+4 \right) {g}^{-n} + \big[ F ( {E}^{2}n{r}^{4}\nonumber \\
& -E{r}^{2}n-2 ) {g}^{n}+2\,g \lbrace {r}^{4} \left( {n}^{2}-\frac{3}{2}n+1 \right)  \left( n+1 \right) {E}^{2}\nonumber \\
& +{r}^{2} \left( \frac{5}{2}{n}^{2}-1 \right) E+\frac{5}{2}n-2 \rbrace  \big] F\Bigg]\Bigg/\Bigg[ 2F\big[\frac{1}{2}{F}^{2}{g}^{2\,n}E{r}^{2}\nonumber \\
& +F \left\lbrace\frac{5}{2}+{r}^{4} \left( {n}^{2}-\frac{3}{2}n +\frac{3}{2} \right) {E}^{2}+\frac{3}{2} \left( n+\frac{4}{3}\right) {r}^{2}E \right\rbrace {g
}^{n}\nonumber\\
& \hspace{-1.5cm}-{g}^{2} \left\lbrace \frac{5}{2}+ \left( -\frac{1}{2}+n \right) {r}^{2}E \right\rbrace
 \left( n-2 \right)  \big] \Bigg].
\end{eqnarray}

\begin{figure*}[t]
\centering
    \includegraphics[width=6cm]{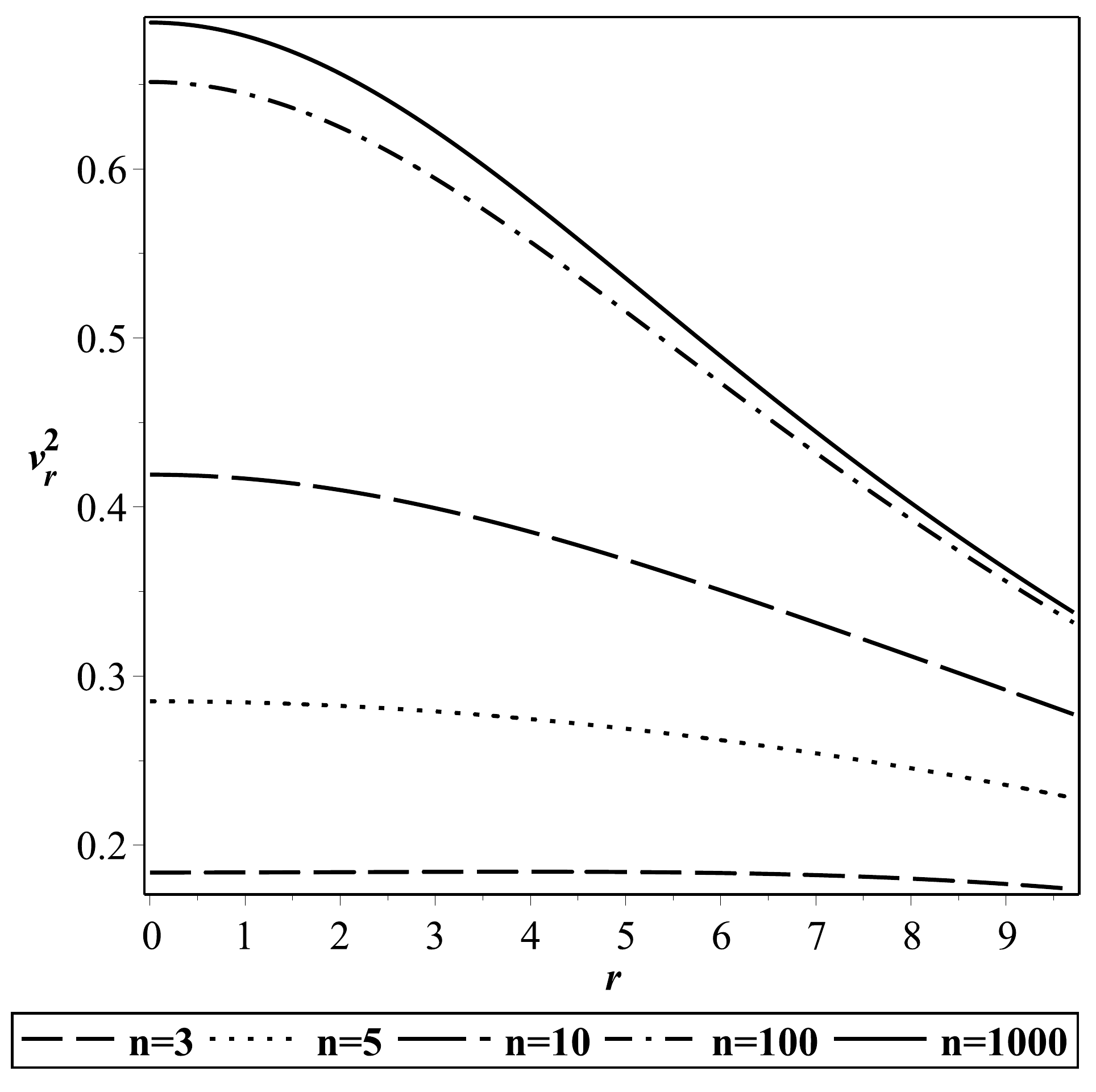}
    \includegraphics[width=6cm]{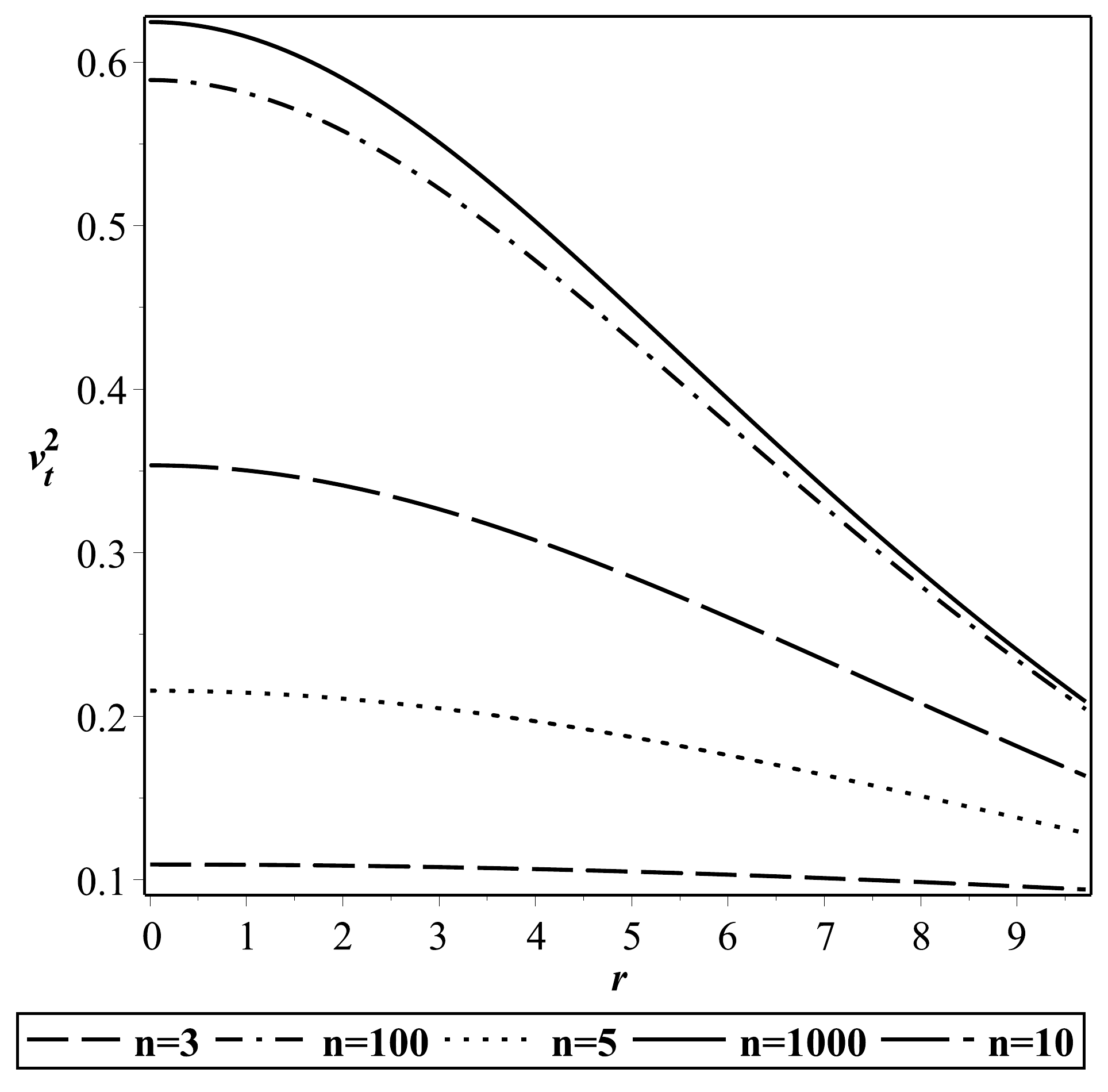}
\caption{Variation of $v_r^2$ (left panel) and $v_t^2$ (right panel) with the radial coordinate $r/R$ for $LMC\,X-4$.} \label{Fig7}
\end{figure*}

\begin{figure*}[!htp]
\centering
    \includegraphics[width=6cm]{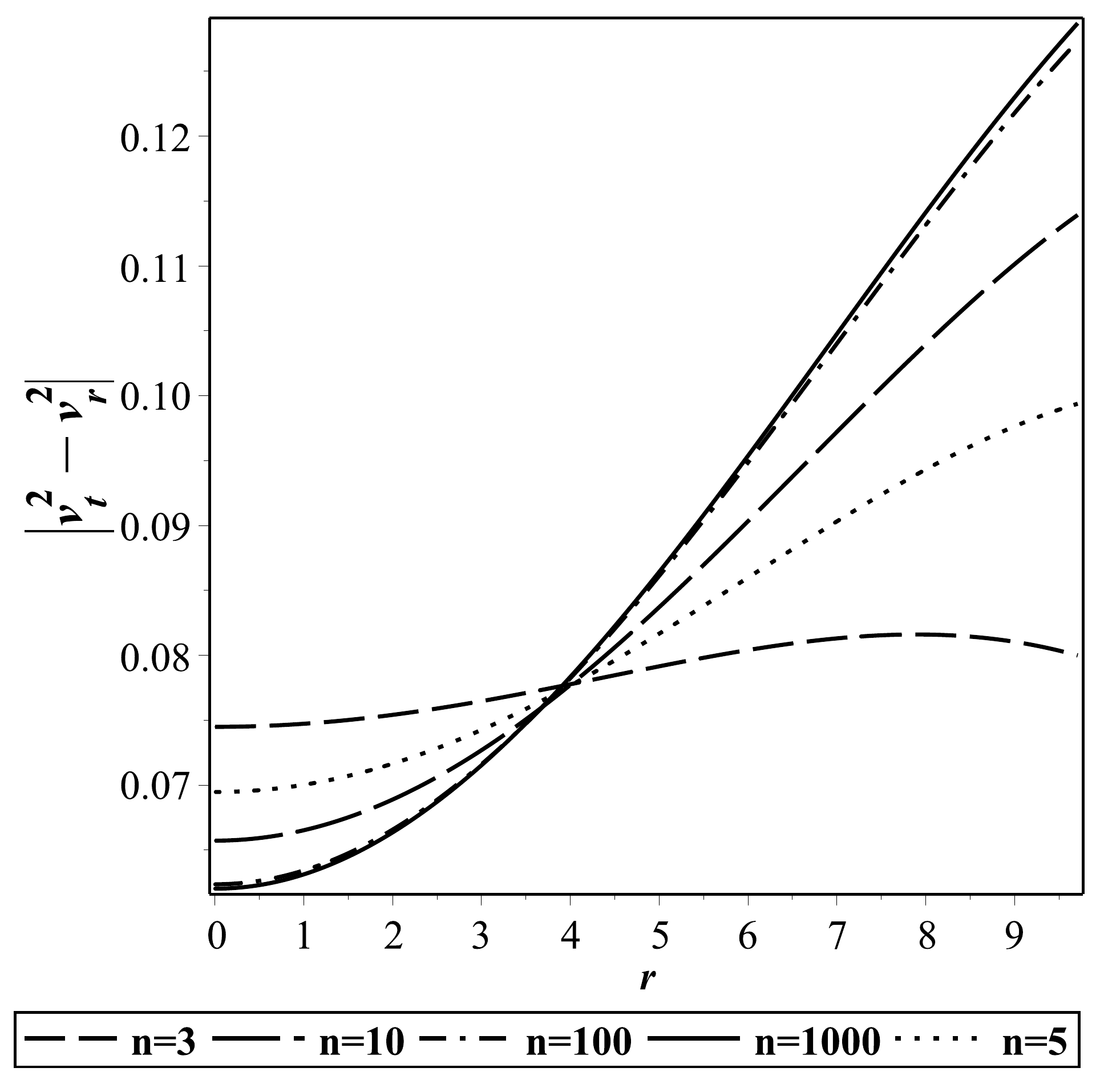}
    \caption{Variation of $v_r^2-v_t^2$ with the radial coordinate $r/R$ for $LMC\,X-4$.} \label{Fig8}
\end{figure*}

From Figs. \ref{Fig7} and \ref{Fig8} we have shown that both the conditions for stability are achieved for
our model.

\subsubsection{Adiabatic index}\label{subsub6.2.3}
The study of the parameter adiabatic index for a compact stellar system is important as it characterizes 
the stiffness of the equation of state for a given density, which also
describes the stability of a compact star $($both relativistic or non
relativistic$)$. In order to have a stable Newtonian compact star,
 \citet{Heintzmann1975} suggest that the adiabatic index must
 exceeds $\frac{4}{3}$ in all the interior points of the stars. But this
 condition modifies for relativistic compact stellar model. For our model we
 have defined the adiabatic index $\gamma_r$ as follows
\begin{equation}
\gamma_r=\left(\frac{p_r+\rho}{p_r}\right) \left[\frac{dp_r}{d\rho}\right]=\left(\frac{p_r+\rho}{p_r}\right)v^2_r. \label{6.2.3.1}
\end{equation}

\begin{figure*}[!htp]
\centering
    \includegraphics[width=6cm]{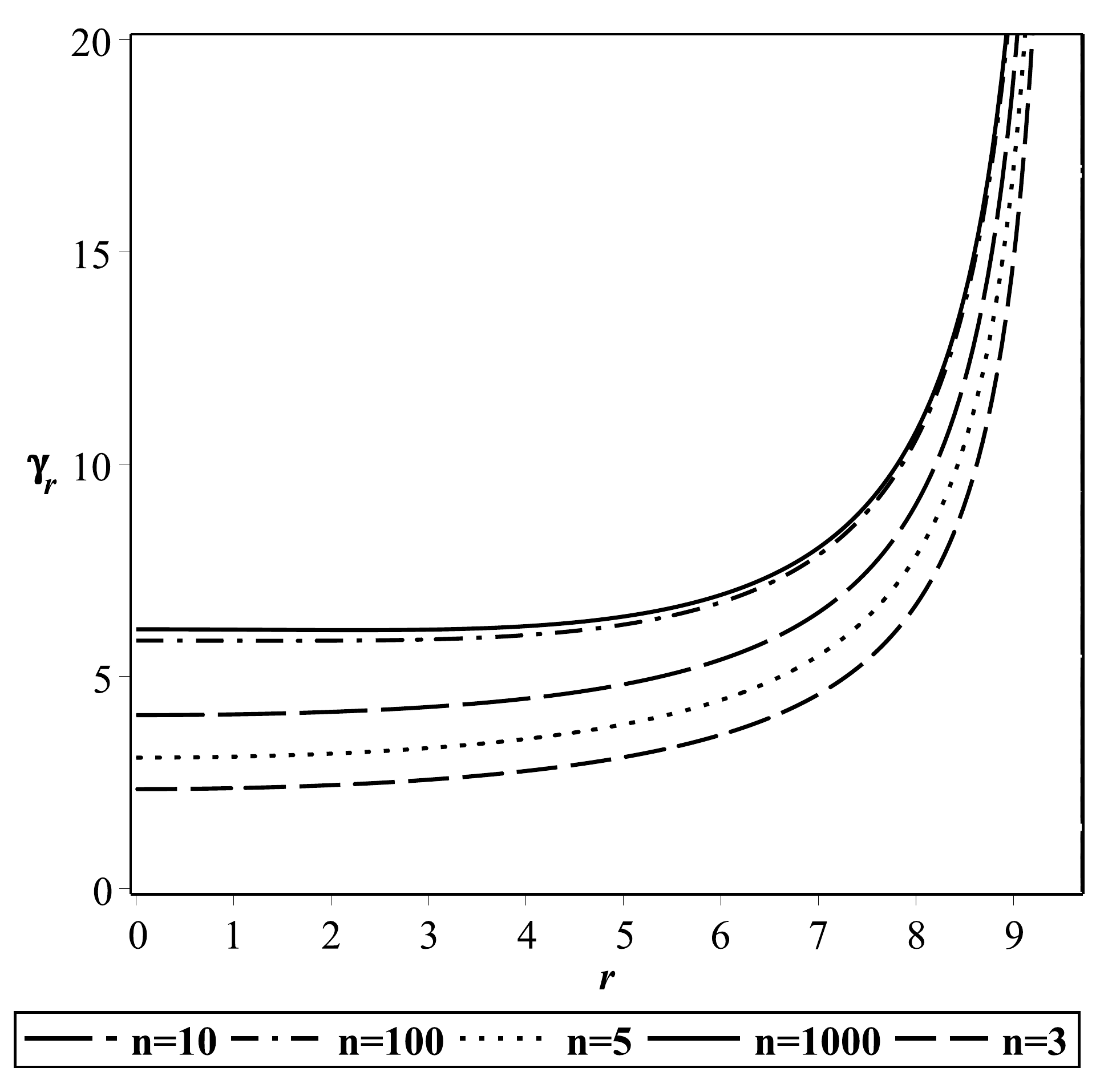}
    \caption{Variation of adiabatic index $(\Gamma_r$ with the radial coordinate $r/R$ for $LMC\,X-4$.} \label{Fig9}
\end{figure*}

The variation of $\gamma_r$ with the fractional radial coordinate are shown in Fig.~\ref{Fig9}. It can be observed that the adiabatic index does satisfy the stability condition (i.e. $\gamma_r>\frac{4}{3}$).

\section{Compactness and Redshift}\label{sec7}
The ratio between mass and radius is defined as the compactification
factor, $u(r)$. The compactification factor plays an important role for
the formation of the core of the compact star and describes how relativistic a star is.
To calculate $u(r)$ we found the mass of a compact star as
\begin{eqnarray}\label{7.1}
m(r)=4\pi \int_{0}^{r}\!\rho \left( r \right) {r}^{2}{\rm d}r={\frac {A{r}^{3}FE}{2FE{r}^{2}+2 \left( E{r}^{2}+1
 \right) ^{2-n}}}.\nonumber \\ 
\end{eqnarray}

From the above Eq. (\ref{7.1}), it can be observed that the mass function is regular within the star and vanishes at the centre, i.e. at $r=0$.

Now the compactification factor, $u(r)$, can be obtained as
\begin{equation}\label{7.2}
u\left(r\right)={\frac {m(r)}{r}}={\frac {A{r}^{2}FE}{2FE{r}^{2}+2 \left( E{r}^{2}+1
 \right) ^{2-n}}}.
\end{equation}

\begin{figure*}[!htp]
\centering
    \includegraphics[width=6cm]{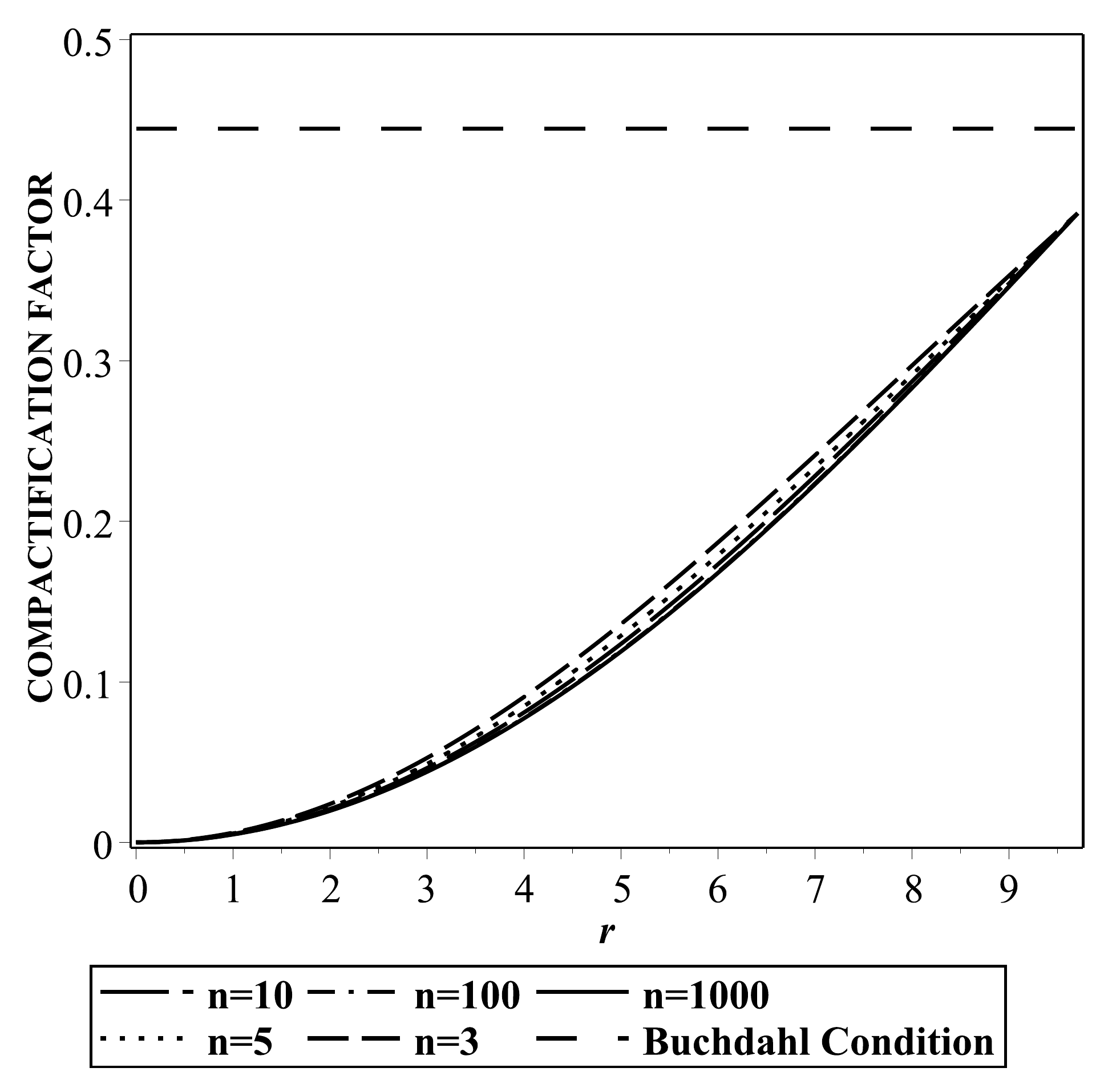}
    \includegraphics[width=6cm]{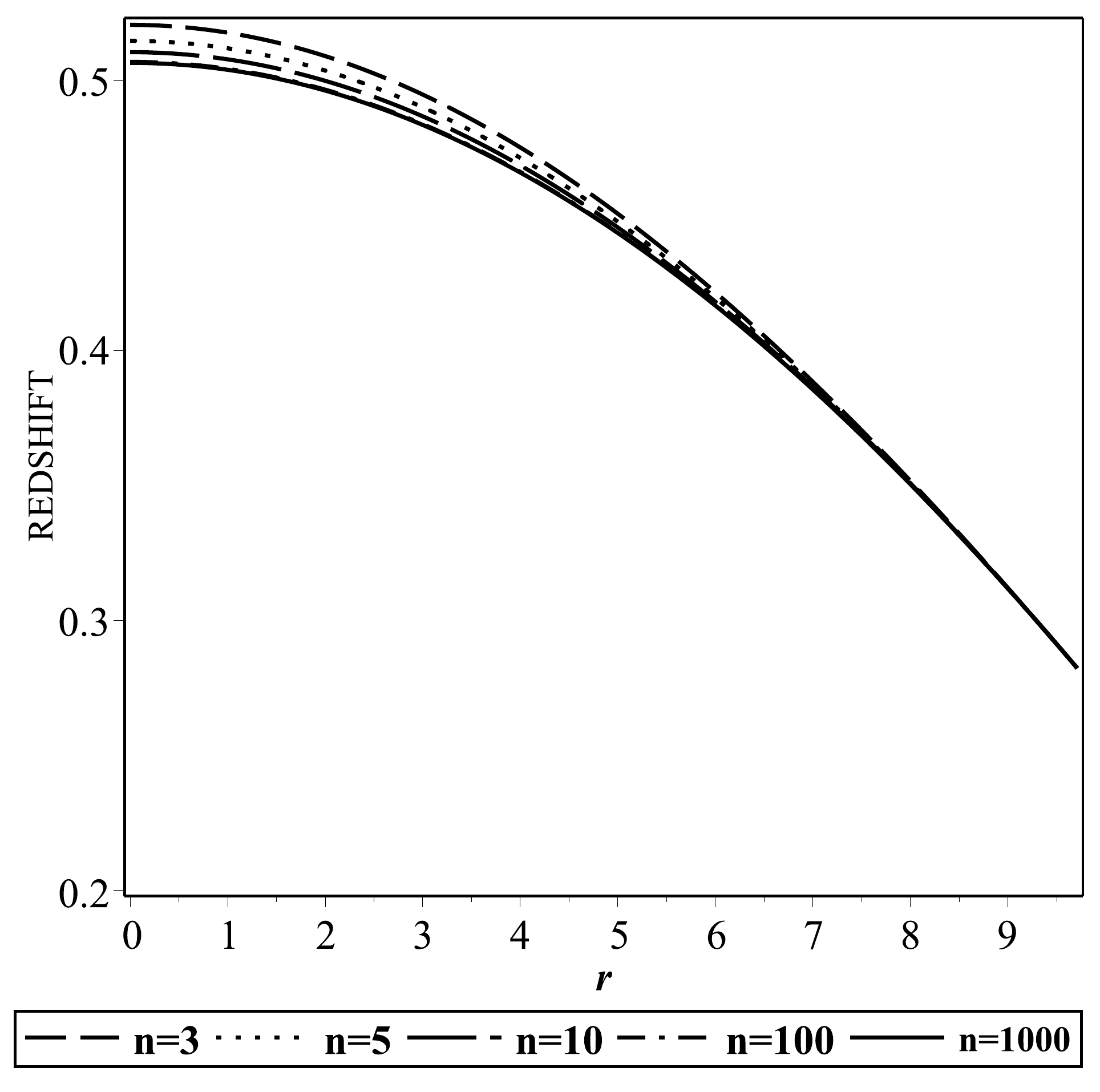}
\caption{Variation of the compactification (left panel) and surface redshift (right panel) with the radial coordinate $r/R$ for $LMC\,X-4$.} \label{Fig10}
\end{figure*}

The variation of the compactification factor with the fractional coordinate is shown in Fig.~\ref{Fig10}~(left panel).
According to \citet{Buchdahl1959} for any physically valid model we have $2u(R)>\frac{8}{9}$. It has been observed that the values of  $2u(R)=\frac{2m(R)}{R}$ for all $n$ satisfies the above condition.

Now the surface redshift of the system is
\begin{equation}\label{7.3}
Z_{{s}}={\frac {1}{\sqrt {P \left( E{r}^{2}+1 \right) ^{n}}}}-1.
\end{equation}
The variation of $Z_{{s}}$ is shown in Fig.~\ref{Fig10} (right panel).

\section{Discussion and conclusion}\label{sec8}
In the present article we have studied a generalized model for static and anisotropic compact stars by reducing a spherically symmetric metric of class 2 into a metric of class 1 by using the Karmarkar condition~\citep{karmarkar1948} in the background of $f(T)$-modified gravity. To this end considering a particular form of metric function $e^{\nu}=P(1+Er^{2})^{n}$ as presented by~\citet{Lake2003} we have determined the metric function $e^{\lambda}$ by employing the above mnetioned Karmarkar condition. Further, we have solved the EFE for the compact stars and demonstrated our calculations considering the stars $LMC~X-4$, $Cen~X-3$~and~$SMC~X-1$ as the representative of compact stars with the suitable choice of the parametric values of $n$. From the nature of the physical parameters as featured in Figs.~\ref{Fig1} to \ref{Fig3}, it is clear that the obtained solution is free from singularity. In the present investigation we have considered a simple linearised form of $f(T)$ given as $f(T)=AT+B$, where $A$ and $B$ are constants.

\begin{table*}
  \centering
	\caption{Numerical values of the parameters $P, E, C$ and $F$ with different values of $n$ for $LMC\,X-4$ where it's mass and radius are respectively $M=1.29~M_{\odot}$~\citep{dey2013} and $R=9.711~km$ and $B=0$ }\label{Table 1}
\scalebox{0.9}{
\begin{tabular}{ ccccccccccccccccccccccccccc}
\hhline{=========}
Values of n & P & E & C & F \\ 
\hline
3 & 0.4325 & $1.276\times{{10}^{-3}}$ & -240.65 & 4.781 \\ 
\
5 & 0.4359 & $7.304\times{{10}^{-4}}$ &  -240.65 & 7.661 \\ 
\
10 & 0.4383 & $3.530\times{{10}^{-4}}$ & -240.65 & 14.895 \\ 
\
100 & 0.4404 & $3.427\times{{10}^{-5}}$ & -240.65 & 145.306 \\ 
\
1000 & 0.4406 & $3.418\times{{10}^{-6}}$ & -240.65 & 1449.518 \\ 
\hhline{=========} 
\end{tabular}  }
  \end{table*}


We get the values of different constants, like $P, E, K, C$ and $B$ after matching the interior metric of class 1 with the exterior Schwarzschild’s exterior metric of class 1. We obtained the value of $B$ as zero for our system. We find from our model that for $A=1$ modified $f(T)$ gravity turns into GR and to avoid that situation we have chosen $A=2$. In our study we have considered variation of $n=3$ to $n=1000$. We find that bellow the value $n=3$ we are not getting any physically acceptable solution. However, one may note that as value of $n$ approaches infinity the metric potential takes form as ${e^{\nu}}=P{e^{{C_1}{r^2}}}$, where ${C_1}=nE$. The variation of the metric potentials $e^{\nu}$ and $e^{\lambda}$ with respect to the radial coordinate are shown in Fig.~\ref{Fig1}. 

In Figs.~\ref{Fig2} to \ref{Fig3} we have featured variation of the density, radial and tangential pressure respectively with the radial coordinate. We find that for different values of $n$ they are maximum at the centre and monotonically decreases inside the stellar configuration and approaches minimum value at the surface. The variation of anisotropy with the radial coordinate is also shown in Fig.~\ref{Fig4}. We find anisotropy is minimum, i.e., zero at the centre and maximum at the surface as predicted by~\citet{Deb2016}. Behaviour of the radial and tangential EOS are also featured in Fig.~\ref{Fig4.1}. In Fig.~\ref{Fig5} we have shown that the proposed anisotropic model is satisfying all the inequalities given in Eqs.(~\ref{6.1.1})-(\ref{6.1.3}) and hence all the energy conditions are satisfied. To verify the stability of the model we have studied i) generalized TOV equation, ii) Herrera cracking concept and iii) adiabatic index. To study the generalized TOV equation we have shown behaviour of the forces with the radial coordinate $r$ in Fig.~\ref{Fig6} and found that for our system sum of the forces is zero for  each values of $n$. In Figs.~\ref{Fig7} and \ref{Fig8} variation of the radial and tangential sound speeds and their differences are shown respectively. We found that for our system both the Hererra cracking concept and causality condition are satisfied. From Fig~\ref{Fig9} it is clear that $\gamma_r<4/3$ and thus it confirms the dynamic stability of the stellar model against the infinitesimal radial adiabatic perturbation.

\begin{table}[t]
	\centering
	\caption{Physical Parameters of $LMC\,X-4$ with different values of $n$ for $LMC\,X-4$, where it's mass and radius are respectively $M=1.29~M_{\odot}$\citep{dey2013} and 	$R=9.711~km$~\citep{Deb2016a} and $B=0$ }\label{Table 2}
\begin{tabular}{cccccccccccc}
\hhline{=========}
Values of n & Central Density & Surface Density & Central Pressure \\ 
& $({gm}/{{cm}^3})$ & $({gm}/{{cm}^3})$ & $({dyne}/{{cm}^2})$  \\
\hline 
3 & $1.961\times{{10}^{15}}$ & $1.045\times{{10}^{15}}$ & $1.5\times{{10}^{35}}$ \\ 
\
5 & $1.799\times{{10}^{15}}$ & $1.092\times{{10}^{15}}$ & $1.646\times{{10}^{35}}$ \\ 
\ 
10 & $1.691\times{{10}^{15}}$ & $1.127\times{{10}^{15}}$ & $1.736\times{{10}^{35}}$ \\ 
\ 
100 & $1.602\times{{10}^{15}}$ & $1.158\times{{10}^{15}}$ & $1.806\times{{10}^{35}}$ \\ 
\ 
1000 & $1.593\times{{10}^{15}}$ & $1.161\times{{10}^{15}}$ & $1.813\times{{10}^{35}}$ \\ 
\hhline{=========}
\end{tabular}
\end{table}


\begin{table}[!htp]
	\centering
	\caption{Physical Parameters of $Cen\,X-3$ with different values of $n$, where it's mass and radius are respectively $M=1.49~M_{\odot}$\citep{dey2013} and $R=10.136~km$~\citep{Deb2016a} and $B=0$ }\label{Table 3}
\begin{tabular}{cccccccccccc}
\hhline{=========}
Values of n & Central Density & Surface Density & Central Pressure \\ 
& $({gm}/{{cm}^3})$ & $({gm}/{{cm}^3})$ & $({dyne}/{{cm}^2})$  \\
\hline 
3 & $2.091\times{{10}^{15}}$ & $1.030\times{{10}^{15}}$ & $1.967\times{{10}^{35}}$ \\ 
\
5 & $1.887\times{{10}^{15}}$ & $1.083\times{{10}^{15}}$ & $2.122\times{{10}^{35}}$ \\ 
\
10 & $1.754\times{{10}^{15}}$ & $1.122\times{{10}^{15}}$ & $2.212\times{{10}^{35}}$ \\ 
\
100 & $1.646\times{{10}^{15}}$ & $1.157\times{{10}^{15}}$ & $2.277\times{{10}^{35}}$ \\ 
\
1000 & $1.636\times{{10}^{15}}$ & $1.161\times{{10}^{15}}$ & $2.283\times{{10}^{35}}$ \\ 
\hhline{=========}
\end{tabular}
\end{table}

\begin{table}[!htp]
	\centering
	\caption{Physical Parameters of $SMC\,X-1$ with different values of $n$, where it's mass and radius are respectively $M=1.04~M_{\odot}$\citep{dey2013} and $R=9.095~km$~\citep{Deb2016a} and $B=0$ }\label{Table 4}
\begin{tabular}{cccccccccccc}
\hhline{=========}
Values of n & Central Density & Surface Density & Central Pressure \\ 
& $({gm}/{{cm}^3})$ & $({gm}/{{cm}^3})$ & $({dyne}/{{cm}^2})$  \\
\hline 
3 & $1.811\times{{10}^{15}}$ & $1.066\times{{10}^{15}}$ & $1.053\times{{10}^{35}}$ \\ 
\ 
5 & $1.692\times{{10}^{15}}$ & $1.105\times{{10}^{15}}$ & $1.178\times{{10}^{35}}$ \\ 
\
10 & $1.610\times{{10}^{15}}$ & $1.135\times{{10}^{15}}$ & $1.260\times{{10}^{35}}$ \\ 
\
100 & $1.542\times{{10}^{15}}$ & $1.161\times{{10}^{15}}$ & $1.325\times{{10}^{35}}$ \\ 
\
1000 & $1.535\times{{10}^{15}}$ & $1.164\times{{10}^{15}}$ & $1.332\times{{10}^{35}}$ \\ 
\hhline{=========}
\end{tabular}
\end{table}

For the chosen compact star $LMC~X-4$ having mass $1.29~{M_{\odot}}$\citep{dey2013} and radius $9.711~km$~\citep{Deb2016a} we have presented values of the constants with different values of $n$ in Table~\ref{Table 1}. We have also predicted different physical parameters of stellar model, like the central density $({{\rho}_c})$, surface density $({{\rho}_s})$ and central pressure $(p_c)$ for different values of $n$ in Table~\ref{Table 2},~\ref{Table 3}~and`\ref{Table 4} for the compact stars $LMC~X-4$, $Cen~X-3$~and~$SMC~X-4$ respectively. From Table~\ref{Table 2} we find that as the value of $n$ increases the value of the corresponding physical parameter also increases. We have shown behaviour of the redshift and compactification factor with the radial coordinate for different values of $n$ in Fig~\ref{Fig10}.

As a final remark, our proposed stellar model is completely stable and consistent with all the above mentioned physical tests, hence it is suitable to study anisotropic compact stellar system under the framework of $f(T)$-modified gravity.

\section*{Acknowledgments}
SR is thankful to the Inter-University Centre for
Astronomy and Astrophysics (IUCAA), Pune, India and The Institute of
Mathematical Sciences, Chennai, India for providing all types
of working facility and hospitality under the Visiting Associateship programme.
MK was supported by Russian Science Foundation and carried in the framework
of MEPhI Academic Excellence Project (contract 02.a03.21.0005,
27.08.2013). A part of this work was completed while DD was 
visiting the IUCAA, Pune, India and the author gratefully acknowledges the warm hospitality there.

\end{document}